\documentclass[12pt]{article}

\usepackage[utf8]{inputenc}
\usepackage{bbm}
\usepackage{bm}
\usepackage{mathrsfs}
\usepackage{amsfonts}
\usepackage{float}
\usepackage{amsthm, mathtools}
\usepackage{epsfig,rotating}
\usepackage{amssymb,amsmath}
\usepackage{latexsym}
\usepackage{graphicx}
\usepackage{geometry}
\usepackage{caption,subcaption}
\usepackage{multirow}
\usepackage{booktabs,natbib}
\usepackage{comment}
\usepackage{natbib}
\usepackage{graphicx}
\usepackage{xcolor}
\usepackage{diagbox}
\usepackage{comment}
\usepackage{hyperref}
\usepackage[ruled,vlined]{algorithm2e}
\bibpunct{(}{)}{;}{a}{,}{,}

\newtheorem{definition}{\noindent D{\footnotesize EFINITION}}
\newtheorem{assumption}{\noindent A{\footnotesize SSUMPTION}}
\newtheorem{theorem}{\noindent T{\footnotesize HEOREM}}
\newtheorem{proposition}{\noindent P{\footnotesize ROPOSITION}}

\newtheorem{lemma}{\noindent L{\footnotesize EMMA}}

\newcommand{\RN}[1]{%
  \textup{\uppercase\expandafter{\romannumeral#1}}%
}

\newcommand{\Rbb}{\mathbb{R}}

\newcommand{\Bscr}{\mathscr{B}}

\newcommand{\Gcal}{\mathcal{G}}

\newcommand{\Bcal}{\mathcal{B}}

\newcommand{\Hcal}{\mathcal{H}}

\newcommand{\Fcal}{\mathcal{F}}
\newcommand{\Pcal}{\mathcal{P}}

\newcommand{\Scal}{\mathcal{S}}
\newcommand{\Ocal}{\mathcal{O}}

\newcommand{\hA}{\hat{A}}
\newcommand{\hB}{\hat{B}}
\newcommand{\sA}{A^*}
\newcommand{\sB}{B^*}

\newcommand{\indep}{\rotatebox[origin=c]{90}{$\models$}}
\newcommand{\tr}{\mathrm{tr}}
\newcommand{\ran}{\mathrm{ran}}
\newcommand{\E}{\mathrm{E}}
\newcommand{\W}{\mathscr{W}_2(M)}

\newcommand{\la}{\langle}
\newcommand{\ra}{\rangle}
\newcommand{\var}{\mathrm{var}}

\newcommand{\cov}{\mathrm{cov}}

\DeclareMathOperator{\diag}{\mathrm{diag}}

\def\sp{{\mathrm{span}}}

\def\eop{\hfill $\Box$ \\ }

\def\nano{\scriptscriptstyle}
\newcommand\hi[1]{^{#1}}
\def\real{\mathbb R}
\newcommand\ca[1]{{\cal{#1}}}
\newcommand\lo[1]{_{#1}}

\def\Var{\mathrm{Var}}

\def\ran{\mathrm{ran}}

\def\cov{\mathrm{cov}}
\def\nano{\scriptscriptstyle}
\def\inv{\hi{\nano -1}}

\def\nano{\scriptscriptstyle}
\def\of{\mbox{\raisebox{1pt}{$\nano{\circ}$}}}
\def\ka{\kappa}

\def\diag{\mathrm{diag}}

\def\diag{\mbox{diag}}

\newcommand{\trans}{^{\mbox{\tiny {\sf T}}}}

\makeatletter
\def\old@comma{,}
 \catcode`\,=13
 \def,{%
   \ifmmode%
     \old@comma\discretionary{}{}{}%
   \else%
     \old@comma%
   \fi%
 }
 \makeatother

\title{Nonlinear Sufficient Dimension Reduction for Distribution-on-Distribution Regression}
\author{
 Qi Zhang, Bing Li, and Lingzhou Xue\\  Department of Statistics, Pennsylvania State University
}
\date{First Version: December 2021; \\ This Version: February 2023.}
\begin{document}

\maketitle

\begin{abstract}
We introduce a new approach to nonlinear sufficient dimension reduction in cases where both the predictor and the response are distributional data, modeled as members of a metric space. Our key step is to build  universal kernels (cc-universal) on the metric spaces, which results in reproducing kernel Hilbert spaces for the predictor and response that are rich enough to characterize the conditional independence that determines sufficient dimension reduction. For univariate distributions, we construct the universal kernel using the Wasserstein distance, while for multivariate distributions, we resort to the sliced Wasserstein distance. The sliced Wasserstein distance ensures that the metric space possesses similar topological properties to the Wasserstein space, while also offering significant computation benefits. Numerical results based on synthetic data show that our method outperforms possible competing methods. The method is also applied to several data sets, including fertility and mortality data and Calgary temperature data.
\end{abstract}


\section{Introduction}
In modern statistical applications, complex data objects such as random elements in general metric spaces are commonly encountered. However, these data objects do not conform to the operation rules of Hilbert spaces and lack important properties such as inner products and orthogonality, making them difficult to analyze using traditional multivariate and functional data analysis methods.   An important example of the metric space-valued data objects is the distributional data, which can be modeled as random probability measures that satisfy specific regularity conditions. Recently, there has been an increasing interest in this type of data. \cite{petersen2019frechet} extended the classical regression to Fr\'echet regression, making it possible
to handle univariate distribution on scalar or vector regression. \cite{fan2021conditional} extended the Fr\'echet regression framework to the case of multivariate response distributions. Besides scalar or vector-valued predictors, the relationship between two distributions is also becoming increasingly important.  \cite{petersen2016functional} proposed the log quantile density (LQD) transformation to transform the densities of these distributions to unconstrained functions in the Hilbert space $L\lo 2$. \citet{chen2019lqd} further applied function-to-function linear regression to the LQD transformations of distributions and mapped the fitted responses back to the Wasserstein space through the inverse LQD transformation. \cite{chen2020wasserstein} proposed a distribution-on-distribution regression model by adopting the Wasserstein metric and shows that it works better than the transformation methods in \citet{chen2019lqd}.

Distribution-on-distribution regression encounters similar challenges to classical regression, including the need for exploratory data analysis, data visualization, and improved estimation accuracy through data dimension reduction. In classical regression, sufficient dimension reduction (SDR) has proven to be an effective tool for addressing these challenges. To set the stage, we outline the classical sufficient dimension reduction (SDR) framework. Let $X$ be a $p$-dimensional random vector in $\Rbb\hi p$ and $Y$ a random variable in $\Rbb$. Linear SDR  aims to find a subspace $\mathcal{S}$ of $\Rbb\hi p$ such that $ Y\indep X|P\lo \Scal X$, where $P\lo \Scal$ is the projection on to $\Scal$ with respect to the usual inner product in $\Rbb\hi p$. As an extension of linear SDR, \cite{li2011principal} and \cite{lee2013general} propose the general theory of nonlinear sufficient dimension reduction, which  seeks  a set of nonlinear functions $f\lo 1(X),\dots,f\lo d(X)$ in a Hilbert space such that
$
Y\indep X|f\lo 1(X), \dots, f\lo d(X).
$

In the last two decades, the SDR framework has undergone constant evolution to adapt to increasingly complex data structures. Researchers have extended SDR to functional data \citep{ferre2003functional,hsing2009rkhs,li2017nonlinear,li2022sufficient}, tensorial data \citep{li2010dimension,ding2015tensor}, and forecasting with large panel data \citep{fan2017sufficient,yu2020nonparametric,luo2021inverse} Most recently, \citet{ying2020fr}, \citet{zhang2021dimension}, and \citet{dong2022frechet} have developed SDR methods for cases where the response takes values in a metric space while the predictor lies in Euclidean space. 

Let $X$ and $Y$ be random distributions defined on $M\subseteq \Rbb\hi r$, with finite $p$-th moments ($p\ge 1$). We do allow $X$ and $Y$ to be random vectors, but our focus will be on the case where they are distributions. Modelling $X$ and $Y$ as random elements in  metric spaces $(\Omega\lo X, d\lo X)$ and $(\Omega\lo Y, d\lo Y)$, we seek nonlinear functions $f\lo 1,\dots, f\lo d$ defined on $\Omega\lo X$ such that the random measures $Y$ and $X$ are conditionally independent given  $f\lo 1(X),\dots, f\lo d(X)$. In order to guarantee the theoretical properties of the nonlinear SDR methods and facilitate the estimation procedure, we assume $f\lo 1,\dots,f\lo d$ to reside in a reproducing kernel Hilbert Space (RKHS). While the nonlinear SDR problem can be formulated in much the same way as that for multivariate and functional data, the main new element in this theory that still requires substantial effort is the construction of positive definite and universal kernels on $\Omega \lo X$ and $\Omega \lo Y$. These are needed for constructing unbiased and exhaustive estimators for the dimension reduction problem \citep{li2018sufficient}. 
We achieve this purpose with specific choices of the metrics of Wasserstein distance and sliced Wasserstein distance: we will show how to construct positive definite and universal kernels and the RKHS generated from them to achieve nonlinear SDR for distributional data.

While acknowledging the recent independent work of \citet{virta2022sliced}, who proposed a nonlinear SDR method for metric space-valued data, our work has some novel contributions. First, we focus on distributional data and consider a practical setting where only discrete samples from each distribution are available instead of the distributions themselves, while \citet{virta2022sliced} only illustrated the method with torus data, positive definite matrices, and compositional data. Second, we explicitly construct universal kernels over the space of distributions, which results in an RKHS that is rich enough to characterize the conditional independence. In contrast, \citet{virta2022sliced} only assumes that the RKHS is dense in $L^2$ space, but misses verifications.

The rest of the paper is organized as follows. Section \ref{sec: nonlinear_sdr} defines the general framework of nonlinear sufficient dimension reduction for distributional data. Section \ref{sec: construct_rkhs} shows how to construct RKHS on the space of univariate distributions and multivariate distributions, respectively. Section \ref{sec: gsir} proposes the generalized sliced inverse regression methods for distribution data. Section \ref{sec: asymptotic} establishes the convergence rate of the proposed methods for both the fully observed setting and the discretely observed setting. Simulation results are presented in Section \ref{sec: simulation} to show the numerical performances of proposed methods. In Section \ref{sec: real_data}, we analyze two real applications to human mortality \& fertility data and Calgary extreme temperature data, demonstrating the usefulness of our methods. All proofs are presented in Section \ref{proofs}.

\section{Nonlinear SDR for Distributional Data}\label{sec: nonlinear_sdr}

We consider the setting of distribution-on-distribution regression. Let $(\Omega,\Fcal, P)$ be a probability space. Let $M$ be a subset of $\Rbb\hi r$ and $\Bcal(M)$ the Borel $\sigma$-field on $M$.  Let $\Pcal \lo p(M)$ be the set of Borel probability measures on $(M, \Bcal(M))$ that have finite $p$-th moment and that is dominated by  the Lebesgue measure on $\real \hi r$. We let $\Omega_X$ and $\Omega_Y$ be nonempty subsets of $\Pcal \lo p(M)$ equipped with metrics $d_X$ and $d_Y$, respectively. We let  $\mathcal{B}_X$ and $\mathcal{B}_Y$ be the Borel $\sigma$-fields generated by the open sets in the metric spaces $(\Omega_X, d_X)$ and $(\Omega_Y, d_Y)$. Let $(X, Y)$ be a random element mapping from $\Omega$ to $\Omega_X\times \Omega_Y$, measurable with respect to the product $\sigma$-field $\Bcal_X\times\Bcal_Y$. We denote the marginal distributions of $X$ and $Y$ by $P_X$ and $P_Y$, respectively, and the conditional distributions of $Y|X$ and $X|Y$ by $P_{Y |X}$ and $P_{X|Y}$.

Let $\sigma(X)$ be the sub $\sigma$-field in $\Fcal$ generated by $X$, that is, $\sigma(X) = X\inv\Bcal_X$. Following the terminology in \cite{li2018sufficient}, a sub $\sigma$-field $\Gcal$ of $\sigma(X)$ is called a sufficient dimension reduction $\sigma$-field, or simply a sufficient $\sigma$-field, if $Y\indep X| \Gcal$. In other words, $\Gcal$ captures all the regression information of $Y$ on $X$. As shown in \citet{lee2013general}, if the family of conditional probability measures $\{P_{X|Y}(\cdot|y): y \in \Omega \lo Y\}$ is dominated by a $\sigma$-finite measure, then the  intersection of all sufficient $\sigma$-field is still a sufficient $\sigma$-field. This minimal sufficient $\sigma$-field is called the central $\sigma$-field for $Y$ versus $X$, denoted by $\Gcal_{Y|X}$. By definition, the central $\sigma$-field captures all the regression information of $Y$ on $X$ and is the target that we aim to estimate. 

\def\cenclass{{\mathfrak{S} \lo {Y|X}}}

Let $\Hcal\lo X$ be a Hilbert space of real-valued functions defined on $\Omega\lo X$. We convert estimating the central $\sigma$ field into estimating a subspace of $\Hcal\lo X$. Specifically, we assume that the central $\sigma$-field is generated by a finite set of functions $f_1,\dots, f_d$ in $\Hcal \lo X$, which can be expressed as
\begin{equation}\label{eq:nonlinear sdr}
    Y\indep X| f_1(X),\dots, f_d(X).
\end{equation}
For any sub-$\sigma$-field $\Gcal$ of $\sigma(X)$, let $\Hcal\lo X (\Gcal)$ denote  the subspace of $\Hcal\lo X$ spanned by the function $f$ such that $f(X)$ is $\Gcal$-measurable, that is,
\begin{equation}\label{eq: central class}
\Hcal\lo X(\Gcal)=\overline{\mathrm{span}}\{f\in\Hcal\lo X, f(X)\, \text{is measurable}\, \Gcal\}.    
\end{equation}
We define the central class as $\cenclass=\Hcal\lo X (\Gcal_{Y|X})$ following \eqref{eq: central class}. We say that a subspace $\mathfrak{S}$ of $\Hcal\lo X$ is unbiased if it is contained in $\cenclass$ and  consistent if it is equal to $\cenclass$. To recover the central class $\cenclass$ consistently by an extension of Sliced Inverse Regression \citep{li1991sliced}, we need to assume the central $\sigma$-field is  complete \citep{lee2013general}.
\begin{definition}
A sub $\sigma$-field $\Gcal$ of $\sigma(X)$ is complete if, for each function $f$ such that $f(X)$ is $\Gcal$ measurable and $E[f(X)|Y] = 0$ almost surely $P\lo Y$, we have $f(X) = 0$ almost surely $P\lo X$. We say that $\Hcal\lo X(\Gcal)$ is a complete class for $Y$ versus $X$ if $\Gcal$ is complete $\sigma$-field for $Y$ versus $X$.
\end{definition}

Although our theoretical analysis so far does not require $\Hcal\lo X$ and $\ca H \lo Y$ to be RKHS, using an RKHS provides a concrete framework for establishing an unbiased and  consistent estimator. It also builds a  connection between the classical linear SDR and nonlinear SDR in the sense that $f(x)$ can be expressed as the inner product $\langle f,\kappa(\cdot,x)\rangle$, where $\kappa: \Omega \lo X \times \Omega \lo X\to \real$ is the reproducing kernel. This inner product is a nonlinear extension of $\beta\trans X$ in linear SDR. In the next section, We will describe how to construct RKHS for univariate  and multivariate distributions.

\section{Construction of RKHS}\label{sec: construct_rkhs}

A common approach to constructing a reproducing kernel is to use a classical radial basis function $\varphi(|x-c|)$ (such as the Gaussian radial basis kernel) and substitute the Euclidean distance with the distance in the metric space.  However, not every metric can be used in such a way to produce positive definite kernels. We show that metric spaces that are of negative type can yield positive definite kernels with form $\varphi(\|x-c\|)$. Moreover, as will be seen in Proposition \ref{prop:regopt} and the discussion following it, in order to achieve an unbiased and consistent estimation of the central class $\Gcal \lo {Y|X}$, we need the kernels for $\ca H \lo X$ and $\ca H \lo Y$ to be cc-universal (\citealt{micchelli2006universal}). For ease of reference, we use the term "universal" to refer to cc-universal kernels. We select the Wasserstein metric and sliced Wasserstein metric for our work, as they possess the desired properties for constructing universal kernels.

\subsection{Wasserstein kernel for univariate distributions}\label{sec: rkhs_wass}

For probability measures $\mu_1$ and $\mu_2$ in $\ca P \lo p (M)$, the $p$-Wasserstein distance between $\mu_1$ and $\mu_2$ is defined as the solution of the Kantorovich transportation problem \citep{villani2009optimal}:
\begin{equation*}
	W\lo p(\mu\lo 1,\mu\lo 2) = \left(\inf\lo{\gamma\in\Gamma(\mu\lo 1,\mu\lo 2)} \int\lo{M\times M} \|x-y\|\hi p d\gamma(x, y)\right)\hi {1/p},
\end{equation*}
where $\|\cdot\|$ is the Euclidean metric, and $\Gamma(\mu\lo 1,\mu\lo 2)$ is the space of joint probability measures on $(M\times M, \ca B (M) \times \ca B(M))$ with marginals $\mu\lo 1$ and $\mu\lo 2$.
When $M\subseteq \real$, the $p$-Wasserstein distance has the following explicit quantile representation:
\begin{equation*}
	W\lo p(\mu_1,\mu_2)=\left(\int_0^1\left[F_{\mu_1}\inv(s)-F_{\mu_2}\inv(s)\right]^2\,ds\right)^{1/p},
\end{equation*}
where $F\inv\lo{\mu\lo 1}$ and $F\inv\lo{\mu\lo 2}$ denote the quantile functions of $\mu_1$ and $\mu_2$, respectively.  The set $\ca P \lo p (M)$ endowed with the Wasserstein metric $W \lo p$ is called the Wasserstein metric space, and is denoted by $\mathscr{W} \lo p (M)$. \citet[Theorem 4]{kolouri2016sliced} show that Wasserstein space of absolutely continuous univariate distributions can be isometrically embedded in a Hilbert space, and thus the Gaussian RBF kernel is positive definite.

We now turn to universality. \citet[Theorem 3]{Christmann10universalkernels} showed that if $\Omega\lo X$ is compact and can be continuously embedded in a Hilbert space $\Hcal$ by a mapping $\rho$, then for any analytic function $A:\Rbb \to \Rbb$ whose Taylor series at zero has strictly positive coefficients, the function $\kappa(x, x^\prime) = A (\langle\rho(x),\rho(x^\prime)\rangle_\Hcal)$ defines a c-universal kernel on $\Omega_X$.
To accommodate the scenarios of $M = \real$ and $M = \real \hi r$, we need to go beyond compact metric spaces. For this reason, we use a more general definition of universality that does not require the support of the kernel to be compact, called cc-universality \citep{micchelli2006universal,sriperumbudur2010relation,sriperumbudur2011universality}. Let $\ka\lo X:\Omega \lo X \times \Omega \lo X \to \real$ be a positive definite kernel and $\ca H \lo X$ the RKHS generated by $\ka \lo X$. For any compact set $K$, let $\ca H \lo X (K)$ be the RKHS generated by $\{\ka \lo X (\cdot, x): x \in K \}$. Let $C(K)$ be the class of all continuous functions with respect to the topology in $(\Omega \lo X, d\lo X)$ restricted on $K$.

\begin{definition}\label{definition:universal}\citep{micchelli2006universal} We say that $\ka\lo X$ is universal(cc-universal) if, for any compact set $K \subseteq \Omega \lo X$, any  member $f$ of $C(K)$, and any $\epsilon > 0$, there is an $h \in \ca H \lo X (K)$ such that
$
\sup \lo {x \in K} | f (x) - h (x) | < \epsilon.
$
\end{definition}

Let $\kappa\lo {\mathrm G}(x,x^{\prime})=\exp(-\gamma \mathrm{W}\lo 2\hi 2(x,x^{\prime}))$ and $\kappa\lo {\mathrm L}L(x,x^{\prime})=\exp(-\gamma {\mathrm{W}}\lo 2(x,x\hi {\prime}))$. The subscripts $G$ and $L$ here refer to ``Gaussian" and ``Laplacian", respectively. \cite{zhang2021dimension} showed that both $\kappa\lo{\mathrm{G}}$ and $\kappa\lo{\mathrm L}$ on a complete and separable metric space that can be isometrically embedded into a Hilbert space are universal. We note that if $M$ is separable and complete, then so is $\W$ \citep[Proposition 2.2.8, Theorem 2.2.7]{panaretos2020invitation}. Therefore, We have the following proposition that guarantees the construction of universal kernels on (possibly non-compact) $\W$.

\begin{proposition}\label{prop:cha_wars}
If $M\subseteq\Rbb$ is complete, then  $\kappa_G(x,x^{\prime})$ and $\kappa_L(x,x^{\prime})$ are universal kernels on $\W$.
\end{proposition}

By Proposition \ref{prop:cha_wars}, we construct the Hilbert spaces $\Hcal\lo X$ and $\Hcal\lo Y$ as RKHS generated by Gaussian type kernel $\kappa_G$ or Laplacian type kernel $\kappa_L$. Let $L_2(P_X)$ be the class of square-integrable functions of $X$ under $P_X$. Let $\mathfrak B$ be the set of measurable indicator functions on $\W$, that is,
\begin{align*}
    \mathfrak B = \{I\lo B: B\subseteq \W\,\, \text{is measurable}\}.
\end{align*}
By \citet[Theorem 1]{zhang2021dimension}, $\Hcal\lo X$ is dense in $\mathfrak B$, and hence dense in $\sp\{\mathfrak B\}$, which is the space of simple functions. Since $\sp\{\mathfrak B\}$ is dense in $L_2(P_X)$, $\ca H\lo X$ is dense in $L_2(P_X)$.

\subsection{Sliced-Wasserstein kernel for multivariate distributions}
\label{sec: rkhs_sliced_wass}
For multivariate distributions ($M \subseteq \real \hi r$), the sliced $p$-Wasserstein distance is obtained by computing the average Wasserstein distance of the projected univariate distributions along randomly picked directions. Let $\mu\lo 1$ and $\mu \lo 2$ be two measures in $\Pcal\lo p (M)$, where $M\subseteq\Rbb\hi r, \ r>1$. Let $\mathbb{S}\hi {r-1}$ be the unit sphere in $\Rbb\hi r$. For $\theta \in \mathbb{S}\hi {r-1}$, let $T\lo \theta : \Rbb\hi r \to \Rbb$ be the linear transformation $x \to \langle \theta, x\rangle$, where $\langle\cdot,\cdot\rangle$ is the Euclidean inner-product. Let $\mu \lo 1 \of T \lo \theta \inv$ and $\mu \lo 2 \of T \lo \theta \inv$ be the induced measures by the mapping $T \lo \theta$. The sliced $p$-Wasserstein distance between $\mu\lo 1$ and $\mu\lo 2$ is defined by
\begin{equation*}
	\mathrm{SW}\lo p(\mu\lo 1,\mu\lo 2)=\left(\int\lo{\mathbb{S}\hi{r-1}}W\lo p \hi p(\mu\lo 1 \of T\lo\theta\inv, \mu\lo 2 \of T\lo\theta\inv)\, d\theta\right)\hi{1/p}.
\end{equation*}
It can be verified that  $\mathrm{SW} \lo p$ is indeed a metric. We denote the metric space $(\Pcal\lo p(M), \mathrm{SW}\lo p)$ by $\mathscr{SW}\lo p (M)$ and call it the sliced Wasserstein space. It has been shown (for example, \citet{bayraktar2021strong}) that the sliced Wasserstein metric is a weaker metric than the Wasserstein metric, that is,
$\forall \mu\lo 1,\mu\lo 2\in P\lo p(M)$ with $M\subseteq\Rbb\hi r$, $\mathrm{SW}\lo p (\mu\lo 1,\mu \lo 2)\le W\lo p (\mu\lo 1, \mu\lo 2)$. This relation implies two topological properties of the sliced Wasserstein space that are useful to us, which can be derived from the topological properties of $p$-Wasserstein space established in \citet[Proposition 7.1.5]{ambrosio2004gradient}, and \citet[Chapter 2.2]{panaretos2020invitation}.
\begin{proposition}\label{prop: sw_topo}
If $M$ is a subset of $\Rbb\hi r$, then $\mathscr{SW}\lo p(M)$ is complete and separable. Furthermore, if $M\subseteq \Rbb\hi r$ is compact, then $\mathscr{SW}\lo p(M)$ is compact.
\end{proposition}

With $p = 2$, \cite{kolouri2016sliced} show that
the square of sliced Wasserstein distance is conditionally
negative definite and hence that the Gaussian RBF kernel $\exp(-\gamma \mathrm{SW}\lo 2\hi 2(x,x^{\prime}))$ is a positive definite kernel. The next lemma shows that the Gaussian RBF kernel and Laplacian RBF kernel based on the sliced Wasserstein distance are, in fact,  universal kernels.
\begin{lemma}\label{lemma: cha_wars_multi}
If $M\subseteq\Rbb\hi r(r>1)$ is complete, then both
$\kappa_G(x, x^{\prime}) = \exp(-\gamma \mathrm{SW}\lo 2\hi 2(x,x^{\prime}))$ and $\kappa_L(x,x^{\prime}) = \exp(-\gamma \mathrm{SW}\lo 2 (x,x^{\prime}))$
are universal kernels on $\mathscr{SW}\lo 2(M)$. Furthermore, $\ca H\lo X$ and $\ca H\lo Y$ are dense in $L\lo 2(P\lo X)$ and $L\lo 2(P\lo Y)$, respectively.
\end{lemma}

\section{Generalized Sliced Inverse Regression for Distributional Data}\label{sec: gsir}
This section extends the generalized sliced inverse regression (GSIR) \citep{lee2013general} for distributional data. We call this extension to univariate distribution settings as Wasserstein GSIR, or W-GSIR, and to multivariate distribution settings as Sliced-Wasserstein GSIR, or SW-GSIR.
\subsection{Distributional GSIR and the role of universal kernel}
To model the nonlinear relationships between random elements, we introduce the covariance operator in the RKHS, a concept similar to the constructions in \cite{fukumizu2004dimensionality}, \cite{lee2013general}, \cite{li2017nonlinear}, and \citet[Chapter 12.2]{li2018sufficient}. Let $\Hcal_1$ and $\Hcal_2$ be two arbitrary Hilbert spaces, and let $\Bscr(\Hcal_1,\Hcal_2)$ denote the class of bounded linear operators from $\Hcal_1$ to $\Hcal_2$. If $\Hcal_1 = \Hcal_2 = \Hcal$, we use $\Bscr(\Hcal)$ to denote $\Bscr(\Hcal, \Hcal)$.  For any operator $T \in \Bscr(\Hcal_1, \Hcal_2)$, we use $T^*$ to denote the adjoint operator of $T$, $\mathrm{ker}(T)$ to denote the kernel of $T$, $\mathrm{ran}(T)$ to denote the range of $T$, and $\overline{\ran}(T)$ to denote the closure of the range of $T$. Given two members $f$ and $g$ of $\Hcal$, the tensor product $f\otimes g$ is the operator on $\Hcal$ such that $(f\otimes g)h=f\langle g,h\rangle_{\Hcal}$ for all $h\in \Hcal$. It is important to note that the adjoint operator of $f\otimes g$ is $g\otimes f$.

We define $E[\kappa(\cdot,X)]$, the mean element of $X$ in $\Hcal\lo X$, as the unique element in $\Hcal\lo X$ such that
\begin{equation}\label{eq: mean_rkhs}
	\langle f ,E[\kappa(\cdot,X)]\rangle_{\Hcal\lo X} = E\langle f ,\kappa(\cdot,X)\rangle_{\Hcal\lo X}
\end{equation}
for all $f\in \Hcal\lo X$. Define the bounded linear operator $E[\kappa(\cdot,X)\otimes \kappa(\cdot,X)]$, the second-moment
operator of $X$ in $\Hcal\lo X$, as the unique element in $\Bscr(\Hcal\lo X)$ such that, for all $f$ and $g$ in $\Hcal\lo X$,
\begin{equation}\label{eq: cov_rkhs}
	\langle f ,E[\kappa(\cdot,X)\otimes\kappa(\cdot,X)]g\rangle_{\Hcal\lo X} = E\langle f ,(\kappa(\cdot,X)\otimes\kappa(\cdot,x))g\rangle_{\Hcal\lo X}.
\end{equation}
We write $\mu_X=E[\kappa(\cdot,X)]$, $M\lo{XX}=E[\kappa(\cdot,X)\otimes k(\cdot,X)]$. For Gaussian RBF kernel and Laplacian RBF kernel based on Wasserstein distance or sliced-Wasserstein distance, $\kappa(X, X)$ is bounded and $E[\kappa(X, X)]$ is finite. By Cauchy-Schwartz inequality and Jensen's inequality, it is guaranteed that items on the right-hand side of \eqref{eq: mean_rkhs} and \eqref{eq: cov_rkhs} are well-defined. The existence and uniqueness of $\mu_X$ and $M\lo{XX}$ is guaranteed by Riesz’s representation theorem. We then define the covariance operator $\Sigma\lo{XX}$ as $M\lo{XX} -\mu\lo X\otimes\mu\lo X$. Then, for all $f,g\in \Hcal\lo X$, we have $\cov(f(X),g(X))=\langle f,\Sigma\lo {XX}g \rangle\lo{\Hcal\lo X}$. Similarly, we can define $\mu\lo Y\in\Hcal\lo Y$, $\Sigma\lo{YY}\in\Bscr(\Hcal\lo Y)$, $\Sigma\lo{XY}\in\Bscr(\Hcal\lo X,\Hcal\lo Y)$ and $\Sigma\lo{YX}\in\Bscr(\Hcal\lo Y,\Hcal\lo X)$. By definition, both $\Sigma\lo{XX}$ and $\Sigma\lo{YY}$ are self-adjoint, and $\Sigma\lo{XY}^*=\Sigma\lo{YX}$.

To define the regression operators $\Sigma\lo{XX}\inv\Sigma\lo{XY}$ and $\Sigma\lo{YY}\inv\Sigma\lo{YX}$, we make the following assumptions. Similar regularity conditions are assumed in \cite{li2011principal,lee2013general,li2018sufficient}.
\begin{assumption}\label{assum:ker} \quad
\begin{itemize}
	\item[(1)] $\mathrm{ker}(\Sigma\lo{XX}) = \{0\}$ and $\mathrm{ker}(\Sigma\lo{YY}) = \{0\}$.
	\item[(2)] $\ran(\Sigma\lo{XY})\subseteq\ran(\Sigma\lo{XX})$ and $\ran(\Sigma\lo{YX})\subseteq\ran(\Sigma\lo{YY})$.
	\item[(3)] The operators $\Sigma\lo{XX}\inv\Sigma\lo{XY}$ and $\Sigma\lo{YY}\inv\Sigma\lo{YX}$ are compact.
\end{itemize}
\end{assumption}

Condition (1) amounts to resetting the domains of $\Sigma\lo{XX}$ and $\Sigma\lo{YY}$ to  $\mathrm{ker}(\Sigma\lo{XX})\hi{\perp}$ and $\mathrm{ker}(\Sigma\lo{YY})\hi{\perp}$, respectively. This is motivated by the fact that members of $\mathrm{ker}(\Sigma\lo{XX})$ and $\mathrm{ker}(\Sigma\lo{YY})$ are constants almost surely, which are irrelevant when we consider independence. Since $\Sigma\lo{XX}$ and $\Sigma\lo{YY}$ are self adjoint operators, this assumption is equivalent to resetting  $\Hcal\lo X$ to $\overline{\ran} (\Sigma\lo {XX})$ and $\Hcal\lo Y$ to $\overline{\ran}(\Sigma\lo{YY})$, respectively.
Condition (1) also implies that the mappings $\Sigma\lo{XX}$ and $\Sigma\lo{YY}$ are invertible, though, as we will see,  $\Sigma \lo {XX} \inv$ and $\Sigma \lo {YY} \inv$ are unbounded operators.

Condition (2) guarantees that
$
\ran(\Sigma\lo{XY})\subseteq\mathrm{dom}(\Sigma\lo{XX}\inv)=\ran(\Sigma\lo{XX})$ and $\ran(\Sigma\lo{YX})\subseteq\mathrm{dom}(\Sigma\lo{YY}\inv)=\ran(\Sigma\lo{YY}),$ 
which is necessary to define the regression operators $\Sigma\lo{XX}\inv\Sigma\lo{XY}$ and $\Sigma\lo{YY}\inv\Sigma\lo{YX}$. By Proposition 12.5 of \cite{li2018sufficient}, $\ran(\Sigma\lo{YX})\subseteq\overline{\ran}(\Sigma\lo{YY})$ and $\ran(\Sigma\lo{XY})\subseteq\overline{\ran}(\Sigma\lo{XX})$. Thus the above assumption is not very strong.

As interpreted in Section 13.1 of \cite{li2018sufficient}, Condition (3) in  Assumption \ref{assum:ker} is akin to a smoothness condition.
Even though the inverse mappings  $\Sigma\lo{XX}\inv$ and $\Sigma\lo{YY}\inv$ are well defined, since  $\Sigma\lo{XX} $ and $\Sigma\lo{YY} $ are Hilbert Schmidt operators (\cite{fukumizu2007statistical}), these inverses are unbounded operators. However, these unbounded operators never appear  by themselves, but are always accompanied by  operators multiplied from the right. Condition (3) assumes that the composite operators $\Sigma \lo {XX}\inv \Sigma \lo {XY}$ and $\Sigma\lo{YY}\inv\Sigma\lo{YX}$ are compact. This requires, for example,  that  $\Sigma\lo{YY}\inv\Sigma\lo{YX}$  must send all incoming functions into the low-frequency range of the eigenspaces of $\Sigma\lo{YY}$ with relatively large eigenvalues. That is, $\Sigma \lo {YX}$ and $\Sigma \lo {XY}$ are smooth in the  sense that their outputs are low-frequency components of $\Sigma \lo {YY}$ or $\Sigma \lo {XX}$.

With Assumption \ref{assum:ker} and universal kernels $\kappa\lo X$ and $\kappa\lo Y$, we then have that the range of the regression operator $\Sigma\lo{XX}\inv\Sigma\lo{XY}$ is contained in central class $\cenclass$. Furthermore, if the central class $\cenclass$ is also complete, it can be fully covered by the range of $\Sigma\lo{XX}\inv\Sigma\lo{XY}$. The next proposition adapts the main result of Chapter 13 of \cite{li2018sufficient} to the current context. 
\begin{proposition}\label{prop:regopt}
If Assumption \ref{assum:ker} holds,  
$\ca H \lo X$ is dense in $L \lo 2 (P \lo X)$ and  
$\ca H \lo Y$ is dense in $L \lo 2 (P \lo Y)$, then we have
$
    \ran (\Sigma\lo{XX}\inv\Sigma\lo{XY} ) \subseteq \cenclass.
$
If, furthermore,  $\cenclass$ is complete, then we have
$
    \ran (\Sigma\lo{XX}\inv\Sigma\lo{XY} ) = \cenclass.
$
\end{proposition}
The universal kernels $\ka \lo X$ and $\ka \lo Y$ proposed in Section \ref{sec: construct_rkhs}  guarantees that $\Hcal\lo X$ is dense in $L_2(P_X)$ and $L\lo2(P\lo Y)$, respectively. 

\subsection{Estimation for distributional GSIR}\label{sec: estimation}
By Proposition \ref{prop:regopt}, for any invertible operator $A$, we have
$
    \overline{\ran}(\Sigma\lo{XX}\inv\Sigma\lo{XY}A\Sigma\lo{YX}\Sigma\lo{XX}\inv)\subseteq\cenclass.
$
 Two common choices are $A=I$ and $A=\Sigma\lo{YY}\inv$. When we take $A=\Sigma\lo{YY}\inv$, the procedure is a nonlinear parallel of SIR in the sense that we simply replace  the inner product in the Euclidean space by the inner product in the
 RKHS $\Hcal\lo X$. For easy reference, we refer to   the method using $A=I$ as W-GSIR1 or SW-GSIR1 and $A=\Sigma\lo{XY}\inv$ as  W-GSIR2  or  SW-GSIR2. To estimate the space $\overline{\ran}(\Sigma\lo{XX}\inv\Sigma\lo{XY}A\Sigma\lo{YX}\Sigma\lo{XX}\inv)$, we successively   solve the following generalized eigenvalue problem: 
\begin{align*}
    &\text{maximize}\quad \langle f;\Sigma\lo{XY}A\Sigma\lo{YX} f\rangle_{\Hcal\lo X}\\
    &\text{subject to}\quad \langle f;\Sigma\lo{XX}f\rangle_{\Hcal\lo X} = 1; f \perp \mathrm{span}\{f_1,\dots,f_{k-1}\}, \quad\text{for}\,\, k = 1, 2,\dots, d
\end{align*}
where $f_1, \dots, f_{k}$ are the solutions to this constrained optimization problem in the first $k$ steps.

At the sample level, we estimate $\Sigma\lo{XX},\Sigma\lo{YY},\Sigma\lo{XY}$ and $\Sigma\lo{YX}$ by replacing the expectations $E(\cdot)$ with sample moments $E\lo n(\cdot)$ whenever possible. For example, suppose  we are given i.i.d. sample $(X_1,Y_1), \dots, (X_n,Y_n)$ of $(X,Y)$. We estimate $\Sigma\lo{XX}$ by
\begin{align*}
    \hat\Sigma\lo{XX}=E_n[\kappa(\cdot,X)\otimes \kappa(\cdot,X)]-E_n[\kappa(\cdot,X)]\otimes E_n[\kappa(\cdot,X)].
\end{align*}
The sample  estimates $\hat\Sigma\lo{YY}, \hat\Sigma\lo{XY}$ and $\hat\Sigma\lo{YX}$ for  $\Sigma\lo{YY},\Sigma\lo{XY}$ and $\Sigma\lo{YX}$ are similarly defined.
The subspace $\overline{\ran} (\hat\Sigma\lo{XX})$ and $\overline{\ran} (\hat\Sigma\lo{YY})$ are spanned by the sets
$
    \Bscr_X=\{\kappa(\cdot,X_i)-E_n\kappa(\cdot,X):i=1,\dots,n\},$ and
$ \Bscr_Y=\{\kappa(\cdot,Y_i)-E_n\kappa(\cdot,Y):i=1,\dots,n\}$, respectively.
Let $K_X, K_Y$ denote the $n\times n$ matrix whose $(i,j)$-th entry is $\kappa(X_i,X_j), \kappa(Y_i,Y_j)$ respectively, and let $Q$ denote the projection
matrix $I_n -1_n1_n^T/n$. For two Hilbert spaces $\Hcal_1$, $\Hcal_2$ with spanning systems $\Bscr_1$ and $\Bscr_2$,  and a linear operator $A:\Hcal_1\to\Hcal_2$, we use the notation ${}_{\Bscr_2}[A]_{\Bscr_1}$ to represent the coordinate representation of $A$ relative to spanning systems $\Bscr_1$ and $\Bscr_2$. We then have the following coordinate representations of covariance operators:
\begin{align*}
    &{}_{\Bscr_X}{[\hat\Sigma\lo{XX}]}_{\Bscr_X}=n\inv G_X, 
    {}_{\Bscr_Y}{[\hat\Sigma\lo{YX}]}_{\Bscr_X}=n\inv G_X, \\
    &{}_{\Bscr_X}{[\hat\Sigma\lo{XY}]}_{\Bscr_Y}=n\inv G_Y,  {}_{\Bscr_Y}{[\hat\Sigma\lo{YY}]}_{\Bscr_Y}=n\inv G_Y,
\end{align*}
where $G_X=QK_XQ$ and $G_Y=QK_YQ$. The details are referred to Section 12.4 of \cite{li2018sufficient}.

When $A=I_n$, the generalized eigenvalue problem becomes
\begin{align*}
    \max \quad [f]^T_{\Bscr_X}G_XG_YG_X [f]_{\Bscr_X}\quad  \text{subject to} \quad  [f]_{\Bscr_X}G^2_X [f]_{\Bscr_X} = 1.
\end{align*}
Let $v = G\lo X [f]_{\Bscr_X}$.   To avoid overfitting, we solve this equation for $[f]_{\Bscr_X}$  via Tychonoff regularization, that is, $[f]_{\Bscr_X} = (G_X +\eta_XI_n)\inv v$, where $\eta_X$ is a tuning constant. The problem is then transformed into finding eigenvector $v_1,\cdots,v_d$ of the following matrix
\begin{align*}
    \Lambda^{(1)}_{\mathrm{GSIR}} = (G_X +\eta_X I_n)\inv G_XG_YG_X(G_X +\eta_X I_n)\inv,
\end{align*}
and then set $[f_j]_{\Bscr_X}=(G_X +\eta_X I_n)\inv v_j$ for $j=1,\dots,d$. In practice, we use $\eta_X = \varepsilon_X\lambda_{\max}(G_X)$, where $\lambda_{\max}(G_X)$ is the maximum eigenvalue of $G_X$ and $\varepsilon_X$ is a tuning parameter.

For the second choice $A=\hat\Sigma\lo{YY}\inv$, we also use the regularized inverse $(G_Y +\eta_Y I_n)\inv$, leading to the following generalized eigenvalue problem:
\begin{align*}
    \max [f]^T_{\Bscr_X}G_XG_Y(G_Y +\eta_Y I_n)\inv G_X [f]_{\Bscr_X}\quad  \text{subject to} [f]_{\Bscr_X}G^2_X [f]_{\Bscr_X} = 1.
\end{align*}
To solve this problem, we first compute the eigenvectors  $v_1,\cdots,v_d$ of the matrix
\begin{align*}
    \Lambda^{(2)}_{\mathrm{GSIR}} = (G_X +\eta_X I_n)\inv G_XG_Y(G_Y +\eta_Y I_n)\inv G_X(G_X +\eta_X I_n)\inv,
\end{align*}
and then set $[f_j]_{\Bscr_X}=(G_X +\eta_X I_n)\inv v_j$ for $j=1,\dots,d$. 

\textbf{Choice of tuning parameters.}\label{subsec:choic}
 We use the general cross validation criterion \citep{golub1979generalized} to determine the tuning constant $\varepsilon \lo X$:
\begin{align*}
    \mathrm{GCV}_X (\varepsilon_X) =\frac{\|K_Y -K_X(K_X +\varepsilon_X\lambda_{\max}(K_X)I_n)\inv K_Y \|_F^2}{\{\tr[I_n-K_X(K_X +\varepsilon_X\lambda_{\max}(K_X)I_n)\inv]\}^2}.
\end{align*}
The numerator of this criterion  is the prediction error and the denominator
is to control the degree of overfitting. Similarly, the GCV criterion  for $\varepsilon_Y$ is defined as
\begin{align*}
    \mathrm{GCV}_Y (\varepsilon_Y) =\frac{\|K_X -K_Y(K_Y +\varepsilon_Y\lambda_{\max}(K_Y)I_n)\inv K_X \|_F^2}{\{\tr[I_n-K_Y(K_Y +\varepsilon_Y\lambda_{\max}(K_Y)I_n)\inv]\}^2}.
\end{align*}
We minimize the criteria over grid $\{10^{-6}, \dots, 10\inv,1\}$ to find the optimal tuning constants. We choose the parameters $\gamma_X$ and $\gamma_Y$ in the reproducing kernels $\kappa_X$ and $\kappa_Y$ as the fixed quantities $  \gamma_X=1/({2\sigma_X^2})$
and $\gamma_Y=1/({2\sigma_Y^2})$, where $\sigma_X^2=\binom{n}{2}\inv\sum_{i<j}d(X_i,X_j)^2$, $\sigma_Y^2=\binom{n}{2}\inv\sum_{i<j}d(Y_i,Y_j)^2$ and metric $d(\cdot, \cdot)$ is  $W\lo 2(\cdot, \cdot)$ for univariate distributional data and $SW\lo 2(\cdot, \cdot)$ for multivariate distributional data.

\textbf{Order Determination.}\label{subsec:dimdeter}
To determine the dimension $d$ in (\ref{eq:nonlinear sdr}), we use the BIC type criterion in  \cite{li2011principal} and \cite{li2017nonlinear}. Let
$
    G_n(k)=\sum_{i=1}^k\hat\lambda_i-c_0\lambda_1n^{-1/2}\log(n)k,
$
where $\lambda_i$'s are the eigenvalues of the matrix $\Lambda_{\mathrm{GSIR}}$ and $c_0$ is taken to be 2 when $A=I\lo p$ and 4 when $A = \Sigma\lo{YY}\inv$. Then we estimate $d$ by
\begin{align*}
    \hat{d}=\arg\max\{G_n(k):k=0,1,\dots,n\}.
\end{align*}
Recently developed order-determination methods, such as the ladle estimator \citep{luo2016combining}, can also be directly used to estimate $d$.

\section{Asymptotic Analysis}\label{sec: asymptotic}
In this section, we establish the consistency and convergence rates of W-GSIR and SW-GSIR. We focus on the analysis of Type-I GSIR, where the operator $A$ is chosen as the identity map $I$. The techniques we use are also applicable to the analysis of Type-II GSIR. To simplify the exposition, we define $\Lambda = \Sigma\lo{XX}\inv\Sigma\lo{XY}\Sigma\lo{YX}\Sigma\lo{XX}\inv$ and $\hat\Lambda = (\hat{\Sigma}_{XX} +\eta_n I_n)\inv\hat\Sigma\lo{XY}\hat\Sigma\lo{YX}(\hat\Sigma\lo{XX} +\eta_n I_n)\inv$.

\subsection{Convergence rate for fully observed distribution}
If we assume that the data ${(X_i, Y_i)}_{i=1}^n$ are fully observed, we can establish the consistency and convergence rates of W-GSIR and SW-GSIR without fundamental differences from \cite{li2017nonlinear}. To make the paper self-contained, we present the results here without proof.
\begin{proposition}\label{prop: fullobs}
Suppose $\Sigma\lo{XY}=\Sigma^\beta_{XX}S_{XY}$ far some linear operator $S_{XY}:\Hcal\lo X\to\Hcal\lo Y$ where $0 < \beta < 1$. Also, suppose $n^{-1/2} \preceq \eta_n \prec 0$. Then
\begin{enumerate}
    \item If $S_{XY}$ is bounded, then $\|\hat\Lambda-\Lambda\|_{\mathrm{OP}}=\Ocal_p(\eta_n^\beta+\eta_n\inv n^{-1/2})$.
    \item If $S_{XY}$ is Hilbert-Schmidt, then $\|\hat\Lambda - \Lambda\|_{\mathrm{HS}} = \Ocal_p(\eta_n\hi\beta + \eta_n\inv n^{-1/2})$.
\end{enumerate}
\end{proposition}
The condition $\Sigma\lo{XY}=\Sigma^\beta_{XX}S_{XY}$ is a smoothness condition, which implies the range space of $\Sigma\lo{XY}$ be sufficiently focused on the
eigenspaces of the large eigenvalues of $\Sigma\lo{XX}$. The parameter $\beta$ characterizes the degree of ''smoothness" in the relation between $X$ and $Y$, with a larger $\beta$ indicating a stronger smoothness relation.

By a perturbation theory result in Lemma 5.2 of \citet{koltchinskii2000random}, the eigenspaces of $\hat\Lambda$ converge to those of $\Lambda$ at the same rate if the nonzero eigenvalues of $\Lambda$ are distinct. Therefore, as a corollary of Proposition \ref{prop: fullobs}, the W-GSIR and SW-GSIR estimators are consistent with the same convergence rates.

\subsection{Convergence rate for discretely observed distribution}\label{sec: converge_rate_discrete}
In practice, additional challenges arise when the distributions are not fully observed. Instead, we observe i.i.d.  samples for each $(X_i,Y_i)$, where $i=1,\dots,n$, which is called the discretely observed scenario. Suppose we observe $(\{X_{1j}\}_{j=1}^{r_1}, \{Y_{1k}\}_{k=1}^{s_1}),\dots,(\{X_{nj}\}_{j=1}^{r_1}, \{Y_{nk}\}_{k=1}^{s_n})$, where $\{X_{ij}\}_{j=1}^{r_i}$ and $\{Y_{ik}\}_{k=1}^{s_i}$ are independent samples from $X_i$ and $Y_i$,  respectively. Let $\hat{X}\lo i$, $\hat{Y}_i$ be the empirical measures ${r_i}\inv\sum_{j=1}^{r_i}\delta_{X_{ij}}$ ${s_i}\inv\sum_{j=1}^{s_i}\delta_{Y_{ij}}$, where $\delta_a$ is the Dirac measure at $a$. Then we estimate $d(X_i,X_k)$ and $d(Y_i,Y_k)$ by $d(\hat{X}\lo i,\hat{X}\lo k)$ and $d(\hat{Y_i},\hat{Y}_k)$, respectively. For the convenience of analysis, we assume the sample  sizes are the same, that is, $r_1=\dots=r_n=s_1=\cdots=s_n=m$.  It is important to note that there are two layers of randomness in this situation: the first generates independent samples of distributions $(X_i, Y_i)$ for $i=1,\dots,n$, and the second generates independent samples given each pair of distributions $(X_i, Y_i)$. 

To guarantee the consistency of W-GSIR or SW-GSIR, we need to quantify the discrepancy between the estimated and true distributions by the following assumption.
\begin{assumption}\label{assum:emp_true}
For $i = 1, \dots, n$, $  E[d(\hat{X\lo i}, X\lo i)] =\Ocal(\delta_m)$ and $ E[d(\hat{Y\lo i}, Y\lo i)] =\Ocal(\delta_m)$, where $\delta\lo m\to 0$ as $m \to \infty$.
\end{assumption}
Let $\mu$ be $X\lo i$ or $Y\lo i$ for $i =1,\dots, n$ and $\hat \mu$ be the empirical measure of $\mu$ based on $m$ i.i.d samples. The convergence rate of empirical measures in Wasserstein distance on Euclidean spaces has been studied in several works, including \cite{dereich2013constructive}, \cite{boissard2014mean}, \cite{fournier2015rate}, \citet{weed2019sharp}, and \citet{lei2020convergence}. When $M\subset \real$ is compact, \citet{fournier2015rate} showed that $\E[W\lo2(\hat \mu, \mu)]\lesssim m\hi{-1/4}$. However, when $M$ is unbounded, such as $M= \Rbb$, we need concentration assumptions or moment assumptions on the measure $\mu$ to establish the convergence rate. Let $m\lo q(\mu) := \int\lo M|x|\hi q\,d\mu$ be the $q$-th moment of $\mu$. If $m\lo q(\mu) <\infty$ for some $q>2$, the result of \cite{fournier2015rate} implies that $E[W\lo 2(\hat\mu, \mu)] = \Ocal(m\hi{-1/4}+m\hi{-(q-2)/(2q)})$. If $q>4$, then the term $m\hi{-(q-2)/(2q)}$ is dominated by $m\hi{-1/4}$ and can be removed. If $\mu$ is a log-concave measure, then \citet{bobkov2019one} showed a sharper rate that $\E[W\lo2(\hat \mu, \mu)]\lesssim \sqrt{\log m/m}$.

The convergence rate of empirical measures in sliced Wasserstein distance has been investigated by \citet{lin2020projection}, \citet{niles2022estimation}, and \citet{nietert2022statistical}.  \citet{lin2020projection}. When $M$ is compact, the result of \citet{lin2020projection} indicates that $\E[\mathrm{SW}\lo2(\hat \mu, \mu)]\lesssim m\hi{-1/4}$. When $M = \real\hi r$ and $m\lo q(\mu) <\infty$ for some $q>2$, \citet{lin2020projection} established the rate $E[\mathrm{SW}\lo 2(\hat\mu, \mu)] = \Ocal(m\hi{-1/4}+m\hi{-(q-2)/(2q)})$. A sharper rate is shown in \citet{nietert2022statistical} under the log-concave assumption on $\mu$.

To ensure notation consistency, we define
\begin{align*}
    \hat\Sigma\lo{XY}=E_n[\kappa(\cdot,\hat{X})\otimes \kappa(\cdot,\hat{Y})]-E_n[\kappa(\cdot,\hat{X})]\otimes E_n[\kappa(\cdot,\hat{Y})],\\
\tilde\Sigma\lo{XY}=E_n[\kappa(\cdot,X)\otimes \kappa(\cdot,Y)-E_n[\kappa(\cdot,X)]\otimes E_n[\kappa(\cdot,Y)].    
\end{align*} 
We note that $\hat{X}_1,\dots,\hat{X}_n$ are independent but not necessarily  identically distributed. Despite this, we still write the sample average as $E_n(\cdot)$. Similarly, we define $\hat\Sigma\lo{XX}$ and $\hat\Sigma\lo{YY}$ as the sample covariance operators based on the estimated distribution $\hat{X}_n,\dots,\hat{X}_n$ and $\hat{Y}_1,\dots,\hat{Y}_n$. Under Assumption \ref{assum:emp_true}, we have the following lemma showing the convergence rates of covariance operators.
\begin{lemma}\label{lemma:cov}
Under Assumption \ref{assum:emp_true}, if the kernel $\kappa(z,z^\prime)$ is Lipschitz continuous, that is,
$ \sup\lo{z\hi\prime} |\kappa(z\lo 1, z\hi \prime)-\kappa(z\lo 2, z\hi\prime)|<Cd(z\lo 1, z\lo 2), $ for some  $C>0,$ then $\Sigma\lo{XX}$, $\Sigma\lo{YY}$ and $\Sigma\lo{YX}$ are Hilbert-Schmidt operators, and we have
$\|\hat\Sigma\lo{XX}-\Sigma\lo{XX}\|_{HS}=\Ocal_p(\delta_m+n^{-1/2})$, $\|\hat\Sigma\lo{YY}-\Sigma\lo{YY}\|_{HS}=\Ocal_p(\delta_m+n^{-1/2})$, and $\|\hat\Sigma\lo{XY}-\Sigma\lo{XY}\|_{HS}=\Ocal_p(\delta_m+n^{-1/2}).$
\end{lemma}
Based on Lemma \ref{lemma:cov}, we establish the convergence rate of W-GSIR in the following theorem.
\begin{theorem}\label{thm: disobs}
Suppose $\Sigma\lo{XY}=\Sigma^{1+\beta}_{XX}S_{XY}$ for some linear operator $S_{XY}:\Hcal\lo X\to\Hcal\lo Y$, where $0 < \beta \le 1$. Suppose $\delta_m+n^{-1/2} \preceq \eta_n \prec 0$, then
\begin{enumerate}
    \item If $S_{XY}$ is bounded, then $\|\hat\Lambda-\Lambda\|_{\mathrm{OP}}=\Ocal_p(\eta_n^\beta+\eta_n\inv(\delta_m+n^{-1/2}))$.
    \item If $S_{XY}$ is Hilbert-Schmidt, then $\|\hat\Lambda - \Lambda\|_{\mathrm{HS}} = \Ocal_p(\eta_n^\beta + \eta_n\inv(\delta_m+n^{-1/2}))$.
\end{enumerate}
\end{theorem}
The proof is provided in Section \ref{proofs}. The same convergence rate can be established for SW-GSIR. 
\section{Simulation}\label{sec: simulation}
In this section, we evaluate the numerical performances of W-GSIR and SW-GSIR. We consider two scenarios: univariate distribution on univariate distribution regression and multivariate distribution on multivariate distribution regression. In Section 6.4, we compare the performance of W-GSIR and SW-GSIR with the result using functional-GSIR \citep{li2017nonlinear}. The code to reproduce the simulation results can be found at \url{https://github.com/bideliunian/SDR4D2DReg}.
\subsection{Computational Details}
We use the Gaussian RBF kernel to generate the RKHS.
We consider the discretely observed situation described in Section \ref{sec: converge_rate_discrete}. Specifically, let $\hat{X}_i={m}\inv\sum_{j=1}^{m}\delta\lo{X_{ij}}$ be the empirical distributions for $i=1,\dots, n$. When $X$ is univariate distributions,  for $i, k = 1, \ldots, n$, we  estimate $W_2(X_i,X_k)$ and $W\lo2(Y\lo i, Y\lo k)$ by
\begin{equation*}
    W\lo 2(\hat{X}_i,\hat{X}_k)=\left(\frac{1}{m}\sum_{j=1}^m(X_{i(j)}-X_{k(j)})^2\right)^{1/2}\,\,\text{and}\,\, W\lo 2(\hat{Y}_i,\hat{Y}_k)=\left(\frac{1}{m}\sum_{j=1}^m(Y_{i(j)}-Y_{k(j)})^2\right)^{1/2},
\end{equation*}
respectively, where $X_{i(j)}$ are the $j$-th order statistics of $\{X_{ij}\}_{j=1}^m$.  

When $X$ is multivariate distribution supported on $M \subseteq \real \hi r$, we estimate the sliced Wasserstein distance using a standard Monte Carlo method, that is,
\begin{align*}
	SW\lo p(\hat{X}_i,\hat{X}_k) \approx& \Bigl( \frac{1}{L}\sum_{l=1}^L W\lo 2 \hi 2(\hat{X}_i\circ T_ {\theta_l}\inv, \hat{X}_k\circ T_ {\theta_l}\inv)\Bigr)^{1/2}\\
	=&\Bigl[ \frac{1}{L}\sum_{l=1}^L W\lo 2 \hi 2\Bigl( \frac{1}{m}\sum_{j=1}^{m}\delta_{\la \theta\lo l,X_{ij}\ra}, \frac{1}{m}\sum_{j=1}^{m}\delta_{\la \theta\lo l, X_{kj}\ra}\Bigr)\Bigr]^{1/2},
\end{align*}
where $\{\theta\lo l\}_{l=1}^L$ are i.i.d. samples drawn from the uniform distribution on $\mathbb{S}\hi{r-1}$. The number of samples $L$ controls the approximation error: a larger $L$ gives a more accurate approximation but increases the computation cost. In our simulation settings, we set $L=50$.

To evaluate the difference between estimated  and true predictors, we consider two measures. The first one is the RV Coefficient of Multivariate Rank (RVMR) defined below, which is a generalization of Spearman's correlation in the multivariate case. For two samples of random vectors $U_1,\dots,U_n\in\Rbb^r$ and $V_1,\dots,V_n\in\Rbb^s$, let $\Tilde{U}_i,\Tilde{V}_i$ be their multivariate ranks, that is,
\begin{align*}
    \Tilde{U}_i=\frac{1}{n}\sum_{\ell=1}^n\frac{U_\ell-U_i}{\|U_\ell-U_i\|},\,\,\text{and}\,\, \Tilde{V}_i=\frac{1}{n}\sum_{\ell=1}^n\frac{V_\ell-V_i}{\|V_\ell-V_i\|}.
\end{align*}
Then the RVMR between $\{U_1,\dots,U_n\}$ and $\{V_1,\dots,V_n\}$ is defined as the RV coefficient between $\{\Tilde{U}_1,\dots,\Tilde{U}_n\}$ and $\{\Tilde{V}_1,\dots,\Tilde{V}_n\}$:
\begin{align*}
\mathrm{RVMR}_n(U,V)=\frac{\tr(\cov_n(\Tilde{U},\Tilde{V})\cov_n(\Tilde{V},\Tilde{U}))}{\sqrt{\tr(\var_n(\Tilde{U})^2)\tr(\var_n(\Tilde{V})^2)}}.
\end{align*}
The second one is the distance correlation~\citep{szekely2007measuring}, a well-known measure of dependence between two random vectors of arbitrary dimension.

\subsection{univariate distribution-on-distribution regression}
We generate normal distribution $Y$ with mean and variance parameters being random variables dependent on $X$, that is,
\begin{equation}\label{eq:gene}
	Y = N(\mu_Y, \sigma^2_Y),
\end{equation}
where $\mu \lo Y$ and $\sigma_Y>0$ are random variables generated according to the following models: 
\begin{align*}
    &\text{Model \RN{1}-1}:\, \mu_Y|X\sim N(\exp(W^2_2(X, \mu_1))+\exp(W^2_2(X,\mu_2)), 0.2^2); \sigma_Y=1,\\
    &\text{Model \RN{1}-2}:\, \mu_Y|X\sim N(\exp(W^2_2(X, \mu_1)), 0.2^2); \sigma_Y=\mathrm{Gamma}(W^2_2(X,\mu_2),W\lo 2(X, \mu_2)),\\
    &\text{Model \RN{1}-3}:\,  \mu_Y|X\sim N(\exp(H(X, \mu_1)), 0.2^2); \sigma_Y=\exp(H(X,\mu_2)),\\
    &\text{Model \RN{1}-4}:\,  \mu_Y|X\sim N(E(X), 0.2^2); \sigma_Y=\mathrm{Gamma}(\Var(X),\sqrt{\Var(X)}),
\end{align*}
We let $\mu_1=\mathrm{Beta}(2,1)$ and $\mu_2=\mathrm{Beta}(2,3)$ and generate discrete observations from distributional predictors by $\{X_{ij}\}_{j=1}^{m}\overset{iid}{\sim}\mathrm{Beta}(a_i,b_i)$ where $a_i\overset{iid}{\sim}\mathrm{Gamma}(2,\mathrm{rate}=1)$ and $b_i\overset{iid}{\sim}\mathrm{Gamma}(2,\mathrm{rate}=3)$. We note that the Hellinger distance between two Beta distributions $\mu = \mathrm{Beta}(a_1, b_1)$ and $\nu=\mathrm{Beta}(a_2, b_2)$ can be represented explicitly as
\begin{align*}
    H(\mu, \nu) = 1-\int \sqrt{f\lo {\mu}(t)f\lo {\nu}(t)}\,dt = 1-\frac{B((a\lo 1+a\lo 2) /2, (b\lo 1+b\lo 2)/2)}{\sqrt{B(a\lo 1,b\lo 1)B(a\lo 2,b\lo 2)}},
\end{align*}
where $B(\alpha,\beta)$ is the Beta function.

We compute the distances $W\lo 2(X, \mu_1)$ and $W\lo 2(X, \mu_2)$ by the $L_2$-distance between the quantile functions. We set $n=100,200$, $m=50, 100$ and generate $2n$ samples $(\{X_{ij}\}_{j=1}^{m},\{Y_{ij}\} \lo {j=1} \hi m )_{i=1}^{2n}$. We use half of them to train the nonlinear sufficient predictors via W-GSIR, and then evaluate the RVMR and distance correlation between the estimated and true predictors using the rest of the data set. The tuning parameters and the dimensions are determined by the methods described in Section \ref{sec: estimation}. The experiment is repeated 100 times, and averages and standard errors (in parentheses) of the RVMR and Dcor are summarized in Table \ref{table:d2d_result}. The following are the identified true predictors for each model: Model \RN{1}-1 uses $W\lo2(X, \mu\lo1)$, Model \RN{1}-2 uses $(W\lo2(X, \mu\lo1), W\lo2(X, \mu\lo2))$, Model \RN{1}-3 uses $(H(X, \mu\lo1), H(X, \mu\lo2))$, and Model \RN{1}-4 uses $E(X)$ and $\var(X)$.

\begin{figure}[ht!]
	\centering
  \begin{tabular}{@{}c@{}}
         \includegraphics[width=.3\linewidth]{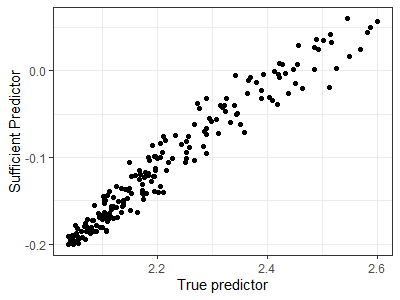} \\[\abovecaptionskip]
    \small (a) 
  \end{tabular}
  \begin{tabular}{@{}c@{}}
         \includegraphics[width=.35\linewidth]{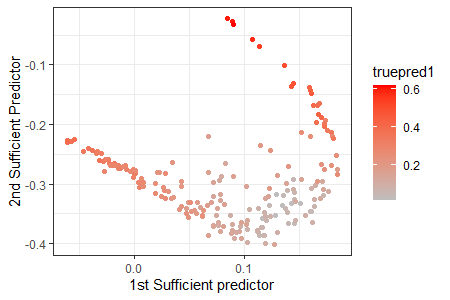} \\[\abovecaptionskip]
    \small (b) 
  \end{tabular}
  \begin{tabular}{@{}c@{}}
         \includegraphics[width=.35\linewidth]{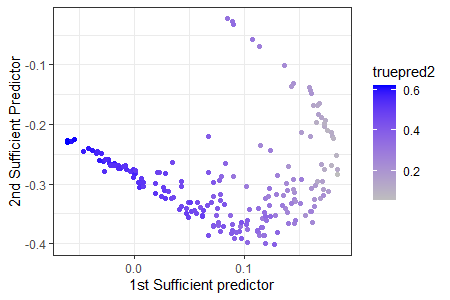} \\[\abovecaptionskip]
    \small (c) 
  \end{tabular}
	\caption{Visualization of W-GSIR1 estimator for (a) Model \RN{1}-1, and (b)(c) Model \RN{1}-2, with $n=200$ and $m=100$. The sufficient predictors are computed via W-GSIR1}\label{fig:wgsir1_d2d_ex1}
\end{figure}

\begin{table}[ht!]
\begin{center}
\begin{tabular}{lccccc}
\hline
& &\multicolumn{2}{c}{W-GSIR1} &\multicolumn{2}{c}{W-GSIR2}\\ \hline
Models& $n\backslash m$  & 50 & 100 & 50 & 100  \\
\hline
&&&RVMR &&\\
\hline
\multirow{2}{*}{\RN{2}-1}
 & 100 & 0.791 ( 0.128 ) & 0.839 ( 0.115 ) & 0.776 ( 0.124 ) & 0.812 ( 0.159 ) \\ 
 & 200 & 0.832 ( 0.091 ) & 0.864 ( 0.087 ) & 0.808 ( 0.114 ) & 0.842 ( 0.129 ) \\
 \cline{2-6}
\multirow{2}{*}{\RN{2}-2}
 & 100 & 0.597 ( 0.187 ) & 0.607 ( 0.206 ) & 0.555 ( 0.236 ) & 0.548 ( 0.235 ) \\ 
 & 200 & 0.694 ( 0.141 ) & 0.681 ( 0.172 ) & 0.709 ( 0.177 ) & 0.688 ( 0.19 ) \\
 \cline{2-6}
\multirow{2}{*}{\RN{2}-3}
 & 100 & 0.846 ( 0.037 ) & 0.880 ( 0.037 ) & 0.836 ( 0.045 ) & 0.859 ( 0.049 ) \\ 
 & 200 & 0.864 ( 0.021 ) & 0.896 ( 0.025 ) & 0.797 ( 0.088 ) & 0.696 ( 0.046 ) \\
 \cline{2-6}
\multirow{2}{*}{\RN{2}-4}
 & 100 & 0.558 ( 0.242 ) & 0.652 ( 0.253 ) & 0.729 ( 0.196 ) & 0.790 ( 0.215 ) \\ 
 & 200 & 0.643 ( 0.221 ) & 0.732 ( 0.183 ) & 0.767 ( 0.169 ) & 0.847 ( 0.145 ) \\ 
\hline
&&&Dcor &&\\
\hline
\multirow{2}{*}{\RN{2}-1}
 & 100 & 0.958 ( 0.024 ) & 0.969 ( 0.022 ) & 0.952 ( 0.029 ) & 0.964 ( 0.034 ) \\ 
 & 200 & 0.967 ( 0.011 ) & 0.974 ( 0.013 ) & 0.963 ( 0.017 ) & 0.970 ( 0.02 ) \\ 
 \cline{2-6}
\multirow{2}{*}{\RN{2}-2}
 & 100 & 0.932 ( 0.037 ) & 0.935 ( 0.041 ) & 0.896 ( 0.071 ) & 0.898 ( 0.066 ) \\ 
 & 200 & 0.952 ( 0.026 ) & 0.948 ( 0.032 ) & 0.934 ( 0.054 ) & 0.932 ( 0.048 ) \\
 \cline{2-6}
\multirow{2}{*}{\RN{2}-3}
 & 100 & 0.971 ( 0.008 ) & 0.978 ( 0.005 ) & 0.968 ( 0.01 ) & 0.974 ( 0.007 ) \\ 
 & 200 & 0.974 ( 0.004 ) & 0.980 ( 0.004 ) & 0.970 ( 0.007 ) & 0.971 ( 0.008 ) \\ 
 \cline{2-6}
\multirow{2}{*}{\RN{2}-4}
 & 100 & 0.921 ( 0.042 ) & 0.936 ( 0.042 ) & 0.937 ( 0.036 ) & 0.947 ( 0.038 ) \\ 
 & 200 & 0.937 ( 0.037 ) & 0.950 ( 0.027 ) & 0.951 ( 0.023 ) & 0.962 ( 0.025 ) \\ 
\hline
\end{tabular}

\end{center}
\caption{RVMR and Distance Correlation with their Monte Carlo standard errors of Scenario \RN{1}}
\label{table:d2d_result}
\end{table}%
Figure \ref{fig:wgsir1_d2d_ex1} (a) displays a scatter plot of the true predictor versus the first estimated sufficient predictor for Model \RN{1}-1. Figures \ref{fig:wgsir1_d2d_ex1} (b) and (c) show the scatter plots of the first two sufficient predictors for Model \RN{1}-2, with the color indicating the values of the true predictor. These figures demonstrate the method's ability to capture nonlinear patterns among predictor random elements.

\subsection{Multivariate distribution-on-distribution regression}
We now consider the scenario where both $X$ and $Y$ are two-dimensional random Gaussian distributions. We generate $Y = N(\mu\lo Y, \Sigma\lo Y)$, where $\mu\lo Y\in \real\hi 2$ and $\Sigma\lo Y\in\real\hi {2\times 2}$ are randomly generated according to the following models:
\begin{itemize}
    \item[] \RN{2}-1: $\mu\lo Y|X = N({{W\lo 2(X, \mu\lo 1)}}(1, 1)\trans, I\lo 2)$ and $\Sigma\lo Y= \diag(1,1).$
    \item[] \RN{2}-2: $\mu\lo Y|X = \sqrt{W\lo 2(X, \mu\lo 1)}(1, 1)\trans$ and $\Sigma\lo Y=\Gamma\Lambda \Gamma\trans$, where $\Gamma = \frac{\sqrt{2}}{2}\begin{pmatrix} 1 & 1\\-1 & 1\end{pmatrix}$, $\Lambda=\diag(|\lambda\lo 1|,|\lambda\lo 2|)$, and $(\lambda\lo 1,\lambda\lo 2)|X\sim N(W\lo 2(X,\mu\lo 2)(1, 1)\trans,0.25I\lo 2)$.
    \item[] \RN{2}-3: $\mu\lo Y|X = N(W\lo 2(X, \mu\lo 1)(1, 1)\trans, I\lo 2)$ and  $\Sigma\lo Y=\Gamma\Lambda \Gamma\trans$, where $\Gamma = \frac{\sqrt{2}}{2}\begin{pmatrix} 1 & 1\\-1 & 1\end{pmatrix}$, $\Lambda=\diag(\lambda\lo 1,\lambda\lo 2)$, and $\lambda\lo 1,\lambda\lo 2|X\overset{i.i.d}{\sim} \mathrm{tGamma}(\mathrm{W}\hi 2 \lo 2(X,\mu_2),\mathrm{W}\lo 2(X, \mu\lo 2), (0.2,2))$.
    \item[] \RN{2}-4: $\mu_Y|X = N(H^2_2(X, \mu_1)(1, 1)\trans, I_2)$ and  $\Sigma_Y=\Gamma\Lambda \Gamma\trans$, where $\Gamma = \frac{\sqrt{2}}{2}\begin{pmatrix} 1 & 1\\1 & -1\end{pmatrix}$, $\Lambda=\diag(\lambda_1,\lambda_2)$,  and $(\lambda\lo 1,\lambda\lo 2)|X\sim \mathrm{tGamma}(H^2(X,\mu_2),H(X, \mu_2), (0.2,2))$.
\end{itemize}
where $\mu\lo1$ and $\mu\lo2$ are two fixed measures defined by 
\begin{align*}
    \mu_1=N((-1,0)\trans, \diag(1,0.5))\,\,\text{and}\,\, \mu_2=N((0,1)\trans, \diag(0.5,1)),
\end{align*}
and $\mbox{tGamma}(\alpha, \beta, (r \lo 1, r \lo 2))$ is the truncated gamma distribution on range $(r\lo 1, r\lo 2)$ with shape parameter $\alpha$ and rate parameter $\beta$. We generate discrete observations of $X\lo i, i=1\, \dots, n$ by $\{X_{ij}\}_{j=1}^{m}\overset{iid}{\sim}{N}(a_i(1,1)\trans,b_iI_2)$ where $a_i\overset{iid}{\sim}{N}(0.5,0.5\hi 2)$ and $b_i\overset{iid}{\sim}{\mathrm{Beta}}(2,3)$. When computing $W\lo2(X, \mu\lo1)$ and $W\lo2(X, \mu\lo2)$, we use the following explicit representations of the Wasserstein distance between two Gaussian distributions:
\begin{align*}
    &W\lo 2\hi 2(N(m\lo 1,\Sigma\lo 1), N(m\lo 2, \Sigma\lo 2)) = \|m\lo 1-m\lo 2\|\hi 2 + \tr \Sigma\lo1 + \tr \Sigma\lo2 - 2\tr\sqrt{\Sigma\lo2\hi{1/2}\Sigma\lo1\Sigma\lo2\hi{1/2}}.
\end{align*}
The following are the identified true predictors for each model: Model \RN{2}-1 uses $W\lo2(X, \mu\lo1)$, Models \RN{2}-2 and \RN{2}-3 uses $(W\lo2(X, \mu\lo1), W\lo2(X, \mu\lo2))$, Model \RN{2}-4 uses $(H(X, \mu\lo1), H(X, \mu\lo2))$.

\begin{figure}[ht!]
	\centering
         \includegraphics[width=0.9\textwidth]{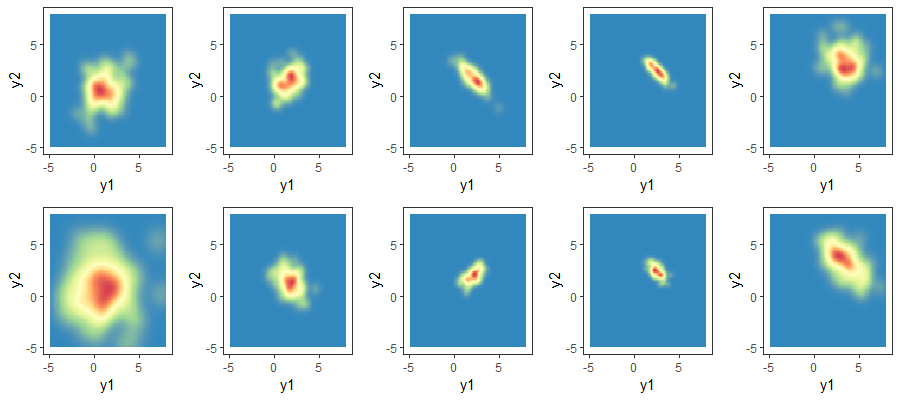}
	\caption{Densities associated with the 10\%, 30\%, 50\%, 70\% and 90\% quantiles (left to right) of estimated predictor (first row) the true predictor (second row) for Model \RN{2}-2 }\label{fig:wgsir1_md2md_ex3}
\end{figure}

Using the true dimensions and the same choices for $n$, $m$ and the tuning parameters, we repeat the experiment 100 times and summarize the average and standard errors of RVMR and distance correlation between the estimated and true predictors in Table \ref{table:md2md_result}.
\begin{table}[ht!]
\begin{center}
\begin{tabular}{lccccc}
\hline
& &\multicolumn{2}{c}{SWGSIR1} &\multicolumn{2}{c}{SWGSIR2}\\ \hline
Models& $n\backslash m$  & 50 & 100 & 50 & 100  \\
\hline
&&& RVMR &&\\
\hline
\multirow{2}{*}{\RN{2}-1}
  & 100 & 0.948 ( 0.063 ) & 0.957 ( 0.049 ) & 0.915 ( 0.13 ) & 0.910 ( 0.15 ) \\ 
  & 200 & 0.958 ( 0.041 ) & 0.970 ( 0.022 ) & 0.921 ( 0.087 ) & 0.934 ( 0.084 ) \\ 
    \cline{2-6}
\multirow{2}{*}{\RN{2}-2}
  & 100 & 0.784 ( 0.036 ) & 0.791 ( 0.033 ) & 0.82 ( 0.038 ) & 0.822 ( 0.036 ) \\ 
  & 200 & 0.783 ( 0.023 ) & 0.791 ( 0.023 ) & 0.834 ( 0.033 ) & 0.824 ( 0.034 ) \\ 
    \cline{2-6}
\multirow{2}{*}{\RN{2}-3} 
  & 100 & 0.744 ( 0.061 ) & 0.755 ( 0.059 ) & 0.806 ( 0.067 ) & 0.812 ( 0.065 ) \\ 
  & 200 & 0.747 ( 0.040 ) & 0.753 ( 0.043 ) & 0.835 ( 0.069 ) & 0.841 ( 0.059 ) \\ 
    \cline{2-6}
\multirow{2}{*}{\RN{2}-4}
  & 100 & 0.499 ( 0.166 ) & 0.500 ( 0.144 ) & 0.57 ( 0.17 ) & 0.567 ( 0.155 ) \\ 
  & 200 & 0.512 ( 0.156 ) & 0.477 ( 0.152 ) & 0.532 ( 0.157 ) & 0.501 ( 0.159 ) \\
  \hline
  &&& Dcor &&\\
  \hline
\multirow{2}{*}{\RN{2}-1}
  & 100 & 0.962 ( 0.024 ) & 0.963 ( 0.025 ) & 0.977 ( 0.018 ) & 0.977 ( 0.021 ) \\ 
  & 200 & 0.963 ( 0.017 ) & 0.964 ( 0.018 ) & 0.973 ( 0.023 ) & 0.97 ( 0.025 ) \\ 
  \cline{2-6}
\multirow{2}{*}{\RN{2}-2} 
  & 100 & 0.967 ( 0.013 ) & 0.967 ( 0.013 ) & 0.973 ( 0.01 ) & 0.975 ( 0.01 ) \\ 
  & 200 & 0.965 ( 0.011 ) & 0.966 ( 0.011 ) & 0.975 ( 0.008 ) & 0.975 ( 0.01 ) \\ 
\cline{2-6}
\multirow{2}{*}{\RN{2}-3}
  & 100 & 0.98 ( 0.009 ) & 0.981 ( 0.008 ) & 0.983 ( 0.007 ) & 0.984 ( 0.006 ) \\ 
  & 200 & 0.979 ( 0.007 ) & 0.979 ( 0.009 ) & 0.982 ( 0.007 ) & 0.983 ( 0.008 ) \\ 
  \cline{2-6}
\multirow{2}{*}{\RN{2}-4}
  & 100 & 0.889 ( 0.031 ) & 0.886 ( 0.036 ) & 0.886 ( 0.033 ) & 0.892 ( 0.03 ) \\ 
  & 200 & 0.893 ( 0.033 ) & 0.886 ( 0.033 ) & 0.887 ( 0.034 ) & 0.889 ( 0.037 ) \\ 
   \hline
\end{tabular}
\end{center}
\caption{RVMR and Distance Correlation with their Monte Carlo standard errors of Scenario \RN{2}. }
\label{table:md2md_result}
\end{table}%
In Figure \ref{fig:wgsir1_md2md_ex3}, we plot the 2-dimensional response densities associated with the 10\%, 30\%, 50\%, 70\%, and 90\% quantiles of estimated predictor (first row) the true predictor (second row) for Model \RN{2}-2. Comparing the plots, we can see that the two-dimensional response distributions show a similar variation pattern, which indicates the method successfully captured the nonlinear predictor in the responses. 
We also see that both the location and scale of the response distribution are captured by the first estimated sufficient predictor. With the increase of the estimated sufficient predictor, the location of the response distribution moves slightly rightward and upward, while the variance of the response distribution decreases at first and then increases.
\subsection{Comparison with functional-GSIR}
Next, we compare the performance of W-GSIR with two methods using the GSIR framework but replacing Wasserstein distance by $L\lo1 $ or $L\lo 2$ distances. We call them $L\lo1$-GSIR and $L\lo2$-GSIR, respectively. Note that $L\lo 2$-GSIR is the same as functional-GSIR (f-GSIR) proposed in \citet{li2017nonlinear}. Theoretically, $L\lo2$-GSIR is an inadequate estimate since an $L \lo 2$ function need not be a density and vice versa. Nevertheless, we still naively implement $L\lo2$-GSIR, treating density curves as $L\lo2$ functions. To make a fair comparison, we first use the Gaussian kernel smoother to estimate the densities based on discrete observations, and then evaluate the $L\lo r$ distances by numerical integration. For $L\lo r$-GSIR ($r=1, 2$), we take the Gaussian type kernel $\kappa(z,z^\prime)=\exp(-\gamma\|z-z^\prime\|_{L\lo r}^2)$, with the same choice of tuning parameters $\gamma$ as described in Subsection \ref{subsec:choic}.
We use  $n=100$, $m=100$,  and repeat the experiment 100 times with $A=I_n$. The results are summarized in Table \ref{table:comp_result}. We see that W-GSIR provides more accurate estimation than both $L\lo1$-GSIR and $L\lo2$-GSIR.
\begin{table}[ht!]
\begin{center}
\begin{tabular}{lcccccc}
\hline
& {$L\lo1$-GSIR1}&{$L\lo2$-GSIR1}&{W-GSIR1}\\
\hline
Models  &&RVMR &&   \\
\hline
\RN{2}-1 & 0.258 ( 0.233 ) & 0.356 ( 0.276 ) & 0.839 ( 0.115 )\\ 
\RN{2}-2 & 0.322 ( 0.236 ) & 0.433 ( 0.244 ) & 0.607 ( 0.206 )\\ 
\RN{2}-3 & 0.307 ( 0.242 ) & 0.359 ( 0.205 ) & 0.880 ( 0.037 )\\ 
\RN{2}-4 & 0.313 ( 0.252 ) & 0.441 ( 0.278 ) & 0.652 ( 0.253)\\ 
\hline
 && Dcor && \\
\hline
\RN{2}-1 & 0.773 ( 0.129 ) & 0.731 ( 0.171 )  & 0.969 ( 0.022 )\\ 
\RN{2}-2 & 0.778 ( 0.173 ) & 0.690 ( 0.203 ) & 0.935 ( 0.041 )\\ 
\RN{2}-3 & 0.779 ( 0.169 ) & 0.688 ( 0.196 ) & 0.978 ( 0.005) \\ 
\RN{2}-4 & 0.799 ( 0.129 ) & 0.740 ( 0.176 ) & 0.936 ( 0.042 )\\
\hline
\end{tabular}
\end{center}
\caption{RVMR and Dcor with Monte Carlo standard errors calculated by $L\lo1$-GSIR, $L\lo2$-GSIR and W-GSIR for Scenarios \RN{1} and \RN{2}}
\label{table:comp_result}
\end{table}%

\section{Applications}\label{sec: real_data}
\subsection{Application to human mortality data}
\begin{figure}[ht!]
	\centering
  \begin{tabular}{@{}c@{}}
         \includegraphics[width=.3\linewidth]{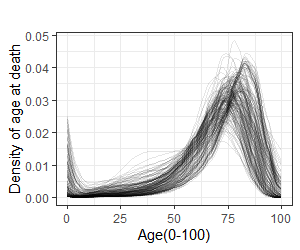} \\[\abovecaptionskip]
    \small (a) 
  \end{tabular}
  \begin{tabular}{@{}c@{}}
         \includegraphics[width=.3\linewidth]{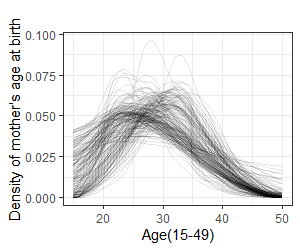} \\[\abovecaptionskip]
    \small (b) 
  \end{tabular}
    \caption{Density of (a) age at death and (b) mother's age at birth for 194 countries.}\label{fig:death_birth_density}
\end{figure}

In this application, we explore the relationship between the distribution of age at death and the distribution of the mother's age at birth. We obtained our data from the UN World Population Prospects 2019 Databases (https://population.un.org), specifically focusing on the years 2015-2020. For each country, we compiled the number of deaths every five years from ages 0-100, and the number of births categorized by mother's age every five years from ages 15-50. We represented this data as histograms with bin widths equal to 5 years. To obtain smooth probability density functions for each country, we used the R package `frechet' to perform smoothing. We then calculated the relative Wasserstein distance between the predictor and response densities. The predictor and response densities are visualized in Figure \ref{fig:death_birth_density}.

\begin{figure}[ht!]
	\centering
  \begin{tabular}{@{}c@{}}
         \includegraphics[width=.24\linewidth]{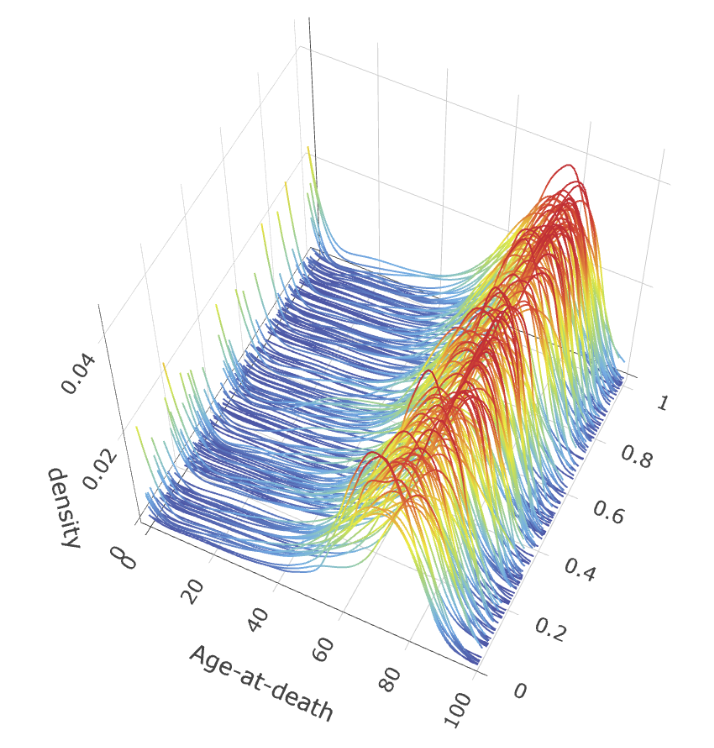} \\[\abovecaptionskip]
    \small (a) 
  \end{tabular}
  \begin{tabular}{@{}c@{}}
         \includegraphics[width=.24\linewidth]{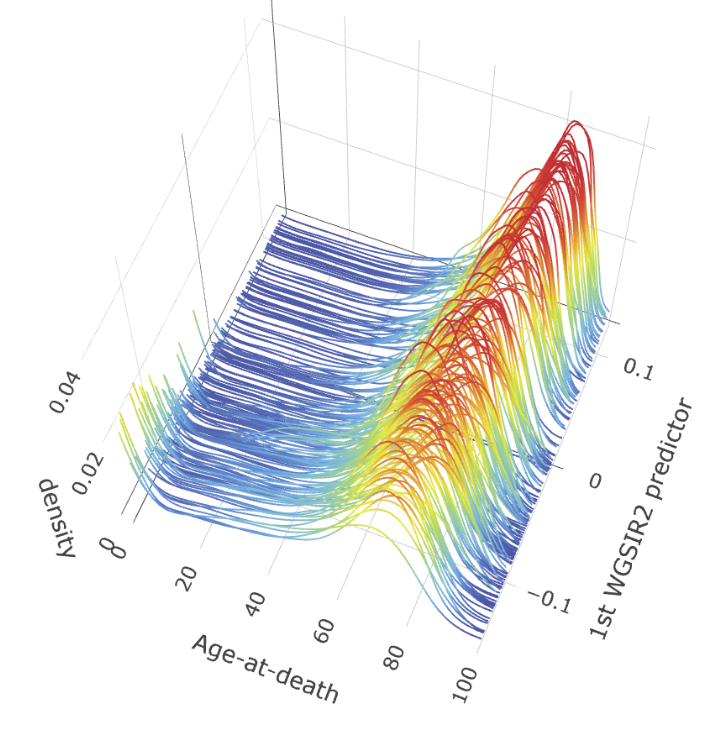} \\[\abovecaptionskip]
    \small (b) 
  \end{tabular}
  \begin{tabular}{@{}c@{}}
         \includegraphics[width=.24\linewidth]{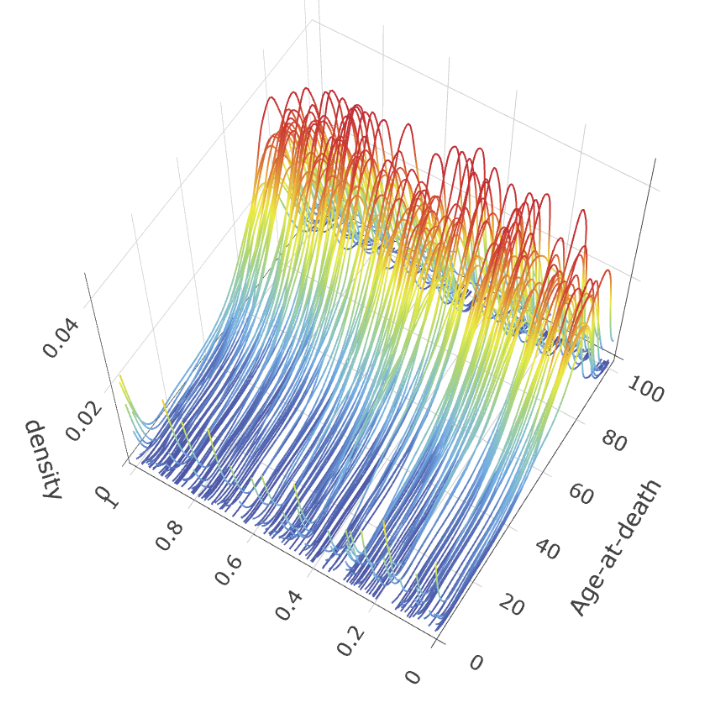} \\[\abovecaptionskip]
    \small (c) 
  \end{tabular}
  \begin{tabular}{@{}c@{}}
         \includegraphics[width=.24\linewidth]{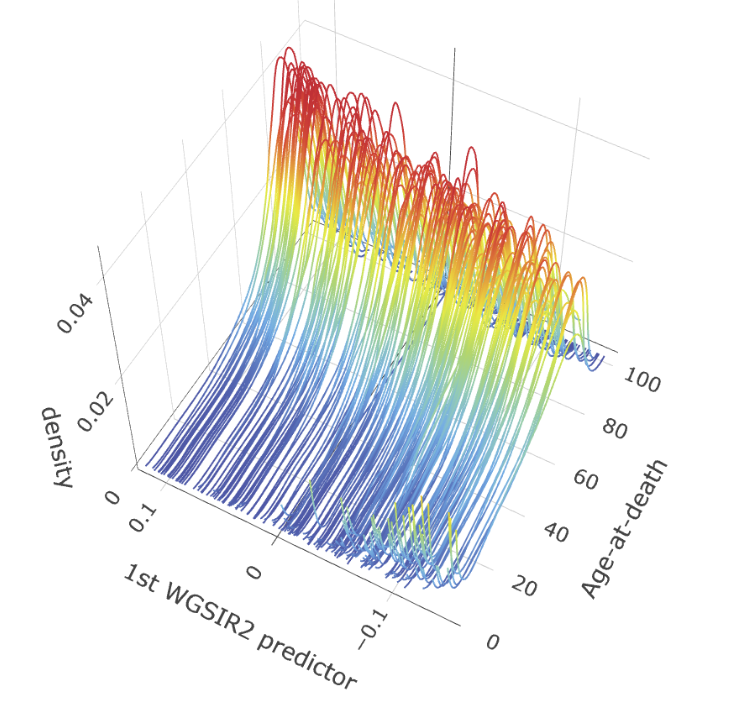} \\[\abovecaptionskip]
    \small (d) 
  \end{tabular}
    \caption{Densities of age at death for 194 countries versus random order in (a) and (c), and versus  the first W-GSIR2 predictor in (b) and (d).}\label{fig:birth_death_wgsir}
\end{figure}

\begin{figure}[ht]
	\centering
    \includegraphics[width=.7\linewidth]{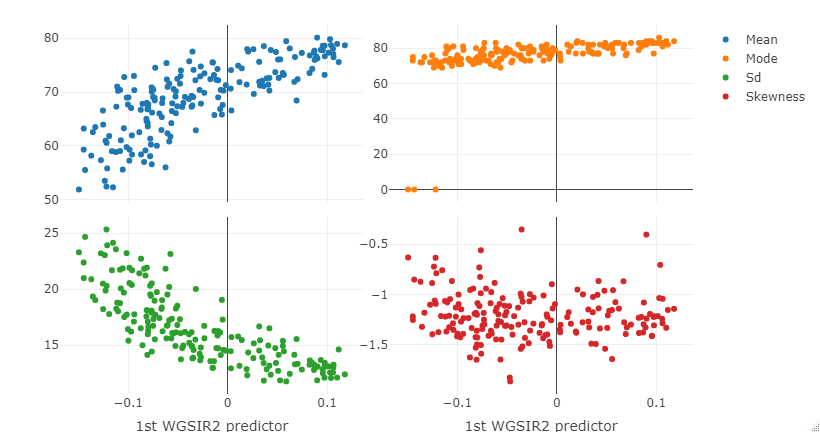} 
    \caption{Summary statistics of mortality distributions for 194 countries versus W-GSIR2 sufficient  predictor.}\label{fig:birth_death_summary}
\end{figure}

We apply the proposed W-GSIR algorithm to the fertility and mortality data. The dimension $d$ of the central class is determined as 1 by the BIC-type procedure described in Subsection \ref{subsec:dimdeter}. We plot the age-at-death distributions  versus the nonlinear sufficient predictors obtained by W-GSIR2 in Figure \ref{fig:birth_death_wgsir}. In Figure \ref{fig:birth_death_summary}, we present the summary statistics of the age-at-death distributions plotted against the sufficient predictors.

Upon examining these plots, we obtained the following insights. The first nonlinear sufficient predictor effectively captures the location and variation information of the mortality distributions. Specifically, as the first sufficient predictor increases, the means of the mortality distributions decrease while the standard deviations increase. This suggests that the population's death age tends to concentrate between 70 and 80 for large sufficient predictor values. Additionally, for densities with small sufficient predictors, there is an uptick at the ends of the 0-age side, which indicates higher infant mortality rates among the countries with such densities.

 \subsection{Application to Calgary temperature data}
\begin{figure}[ht]
	\centering
         \includegraphics[width=\textwidth]{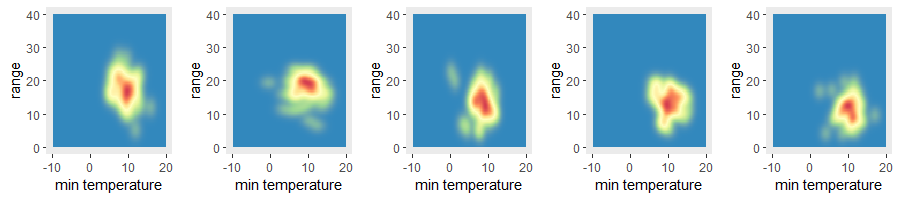}
	\caption{Joint distribution of temperature range and minimum temperature in summer associated with the 10\%, 30\%, 50\%, 70\%, and 90\% quantiles (from left to right) of SWGSIR2 predictor. }\label{fig:spring_summer}
\end{figure}

 In this application, we are interested in the relationship between the extreme daily temperatures in spring (Mar, Apr, and May) and summer (Jun, Jul, and Aug) in Calgary, Alberta.
 We obtained the dataset from \url{https://calgary.weatherstats.ca/}, which contains the minimum and maximum temperatures for each day from 1884 to 2020. These data were previously analyzed in \citet{fan2021conditional}. We focused on the joint distribution of the minimum daily temperature and the difference between the maximum and minimum daily temperatures, which ensures that the distributions have common support. Each pair of daily values was treated as one observation from a two-dimensional distribution, resulting in one realization of the joint distribution for spring and one for summer each year. We then employed the spring extreme temperature distribution to predict the summer extreme temperature distribution. The dataset had $n = 136$ observations, with $m = 92$ discrete values for each joint distribution. We utilized the SW-GSIR method on the data, taking 50 random projections with $\rho\lo X=\rho\lo Y = 1$. The sufficient dimension was determined as 2 using a BIC-type procedure. We illustrated the response summer extreme temperature distributions associated with the five percentiles of the first estimated sufficient predictors in Figure \ref{fig:spring_summer}.
 It is observed from Figure \ref{fig:spring_summer} that as the estimated sufficient predictor value increases, the minimum daily temperature for summer rises slightly while the daily temperature range decreases.

\section{Proofs}\label{proofs}

\subsection{Geometry of Wasserstein Space}
We present  some basic results that characterize $\W$ when $M\subseteq \Rbb$ (i.e., the distributions involved are univariate). Their proofs  can be found, for example, in \cite{ambrosio2004gradient} and \cite{bigot2017geodesic}.
In this case,  $\W$ is a metric space with a formal Riemannian structure \citep{ambrosio2004gradient}.
Let $\mu_0 \in \W$ be a reference measure with a continuous $F_{\mu_0}$. The tangent space at $\mu_0$ is
\begin{align*}
    T_{\mu_0}=\mathrm{cl}_{L_2(\mu_0)}{\{\lambda(F_\mu\inv\circ F_{\mu_0}-\mathrm{id}):\mu\in\W, \lambda > 0\}},
\end{align*}
where, for a set $A\subseteq L_2(\mu_0)$, $\mathrm{cl}_{L_2(\mu_0)}(A)$ denotes the $L_2(\mu_0)$-closure of $A$, and $\mathrm{id}$ is the identity map. The exponential map $\exp_{\mu_0}$ from $T_{\mu_0}$ to $\W$  is defined by $\exp_{\mu_0}({r})= \mu_0 \of ({r}+\mathrm{id}) \inv $, where the right-hand side is the measure on $\W$ induced by the mapping mapping $r + \mathrm{id}$. The logarithmic map $\log_{\mu_0} $ from  $\W$ to $T_{\mu_0}$ is defined by $\log_{\mu_0}(\mu) = F_\mu\inv \circ F_{\mu_0}-\mathrm{id}$. It is known that the exponential map restricted to the image of $\log$ map, denoted as   $\exp_{\mu_0}|_{\log_{\mu_0}(\mu)(\W)}$, is an isometric homeomorphism with inverse $\log\lo {\mu\lo 0}$\citep{bigot2017geodesic}. Therefore, $\log_{\mu_0}$ is a continuous injection from $\W$ to $L_2(\mu_0)$. This embedding guarantees that we can replace the Euclidean distance by the  $\W$-metric in a radial basis kernel to construct a positive definite kernel.

\subsection{Proof of Proposition \ref{prop: sw_topo}}
\begin{proof}
Recall that $\Gamma(\mu\lo 1,\mu\lo 2)$ is the space of joint probability measures on $M\times M$ with marginals $\mu\lo 1$ and $\mu\lo 2$. Let $T\lo \theta \times T\lo \theta$ be the mapping from $M\times M$ to $\Rbb\times \Rbb$ defined by $(T\lo \theta\times T\lo \theta )(x, y) = (T\lo \theta (x), T\lo \theta(y))$. We first show that, if $\gamma\in \Gamma(\mu\lo 1,\mu\lo 2)$, then $\gamma\circ (T\lo\theta\times T\lo\theta)\inv\in \Gamma(\mu\lo 1\circ T\lo\theta \inv, \mu\lo 2\circ T\lo \theta\inv)$. This is true because, for any Borel set $A\subseteq \Rbb$, we have
\begin{align*}
    [\gamma\circ(T\lo\theta\times T\lo\theta)\inv] (A\times T\lo \theta(M)) = \gamma ((T\lo\theta\times T\lo\theta)\inv (A\times T\lo \theta(M)) \\
    = \gamma (T\lo\theta\inv(A)\times M) = \mu\lo 1(T\lo \theta\inv{A}) = \mu\lo 1 \circ T\lo\theta(A),
\end{align*}
and similarly $[\gamma\circ(T\lo\theta\times T\lo\theta)\inv] (T\lo \theta (M) \times A) = \mu\lo 2 \circ T\lo\theta(A)$. Hence, for any $\gamma\in \Gamma(\mu\lo 1,\mu\lo 2)$, we have
\begin{align*}
    W\lo p\hi p(\mu\lo 1\circ T\lo\theta\inv, \mu\lo 2\circ T\lo\theta\inv)
    \le & \int\lo{T\lo\theta(M)\times T\lo\theta(M)} |u-v|\hi p d\gamma\circ(T\lo\theta\times T\lo\theta)\inv(u, v)\\
    = & \int\lo{M\times M} |T\lo\theta (x)-T\lo\theta(y)|\hi p d\gamma(x, y)\\
    \le & \int\lo{M\times M} \|x-y\|\hi p\lo 2 d\gamma(x, y),
\end{align*}
where the last inequality is from the Cauchy-Schwartz inequality.
Therefore,
\begin{align*}
    W\lo p\hi p(\mu\lo 1\circ T\lo\theta\inv, \mu\lo 2\circ T\lo\theta\inv) \le \inf_{\gamma\in \Gamma(\mu\lo 1,\mu\lo 2)} \int\lo{M\times M} \|x-y\|\hi p d\gamma(x, y) = W\lo p \hi p (\mu\lo 1,\mu\lo 2).
\end{align*}
Integrate the left-hand side with respect to $\theta$ and obtain
$
    \mathrm{SW}\lo p (\mu\lo 1, \mu\lo 2)\le W\lo p (\mu\lo 1, \mu\lo 2).
$
Therefore, the $\mbox{SW} \lo p$  distance is a weaker metric than $\mathrm{W}\lo p$ distance, which implies  every open set in $\mathscr{SW}\lo p (M)$ is open in $\mathscr{W}\lo p(M)$. In other words,  $\mathscr{SW}\lo p (M)$ has a coarser topology than $\mathscr{W} \lo p (M) $. Since $M\subseteq\Rbb\hi r$ is separable, so is $\mathscr{W}\lo p(M)$    \citep[Remark 7.1.7]{ambrosio2004gradient}. Therefore,  a countable dense subset of $\mathscr{W}\lo p(M)$ is also a countable dense subset of $\mathscr{SW}\lo p(M)$, implying $\mathscr{SW}\lo p (M)$ is separable. Furthermore, if $M$ is a compact set in $\Rbb\hi r$, then $\mathscr{W}\lo p(M)$ is compact \citep[Proposition 7.1.5] {ambrosio2004gradient}, implying $\mathscr{SW}\lo p (M)$ is compact. This completes the proof  of Proposition \ref{prop: sw_topo}.\eop
\end{proof}
\subsection{Proof of Lemma \ref{lemma: cha_wars_multi}}
\begin{proof}
By Theorem 3.2.2 of \cite{berg1984harmonic}, the kernel $\exp(-\gamma \mathrm{SW}\lo 2\hi 2(x, x\hi \prime))$ is positive definite for
all $\gamma > 0$ if and only if $\mathrm{SW}\lo 2\hi 2(\cdot,\cdot)$ is conditionally negative definite. That is, for any $c\lo 1, \dots, c\lo m \in\Rbb$ with $\sum\lo{i = 1}\hi m c\lo i = 0$, and $x\lo 1, \dots, x\lo m\in\Omega\lo X$, $\sum_{i=1}^m\sum_{j=1}^mc\lo ic\lo j\mathrm{SW}\lo 2\hi 2(x\lo i, x\lo j) \le 0$. \citet[Theorem 5]{kolouri2016sliced} showed the conditional negativity of the sliced Wasserstein distance, which is implied by the negative type of the Wasserstein distance. By \cite{schoenberg1937certain,schoenberg1938metric}, a metric is of negative type is equivalent to the statement that there is a Hilbert space $\Hcal$ and a map $\phi: \mathscr{SW}\lo 2(M) \to \Hcal$ such that $\forall x,x\hi\prime \in \mathscr{SW}\lo 2(M),  \mathrm{SW}\lo 2\hi 2(x,x\hi \prime) = \|\phi(x) - \phi(x\hi\prime)\|\hi 2$. By Proposition \ref{prop: sw_topo}, $\mathscr{SW}\lo 2(M)$ is a complete and separable space. Then by the construction of the Hilbert space, $\Hcal$ is complete and separable. Therefore, there exists a continuous mapping from metric space $\mathscr{SW}\lo 2(M)$ to a complete and separable Hilbert space $\Hcal$. Then by \citet[Theorem 1]{zhang2021dimension}, the Gaussian type kernel $\exp(-\gamma \mathrm{SW}\lo 2\hi 2(x,x^{\prime}))$ is universal. Hence, $\ca H\lo X$ and $\ca H\lo Y$ are dense in $L\lo 2(P\lo X)$ and $L\lo 2(P\lo Y)$, respectively. Same proof applies to the Laplacian-type kernel $\exp(-\gamma \mathrm{SW}\lo 2 (x,x^{\prime}))$. This completes the proof  of Lemma \ref{lemma: cha_wars_multi}.
\eop
\end{proof}

\subsection{Proof of Lemma \ref{lemma:cov}}
\begin{proof}
We will only show the details of the  proof   for the convergence rate of $\|\hat\Sigma\lo{XY}-\Sigma\lo{XY}\|_{\mathrm{HS}}$. By the triangular inequality,  
\begin{equation*}
    \|\hat\Sigma\lo{XY}-\Sigma\lo{XY}\|\lo {\mathrm{HS}}\le\|\hat\Sigma\lo{XY}-\tilde\Sigma\lo{XY}\|\lo {\mathrm{HS}}+\|\tilde\Sigma\lo{XY}-\Sigma\lo{XY}\|\lo {\mathrm{HS}}.
\end{equation*}
By Lemma 5 of \cite{fukumizu2007statistical}, under the assumption that $E[\kappa(X,X)]<\infty$ and $E[\kappa(Y,Y)]<\infty$, we have
\begin{equation}
    E\|\tilde\Sigma\lo{XY}-\Sigma\lo{XY}\|\lo {\mathrm{HS}}=\Ocal(n^{-1/2}).\tag{S.1}\label{eq:lemmacov0}
\end{equation}
Now, we derive a convergence rate for $\|\hat\Sigma\lo{XY}-\tilde\Sigma\lo{XY}\|\lo {\mathrm{HS}}$. For simplicity,  let  $\hat{F}_i=\kappa(\cdot,\hat X_i)$, $\tilde{F}_i=\kappa(\cdot,X_i)$, $\hat{G}_i=\kappa(\cdot,\hat Y_i)$, and  $\tilde{G}_i=\kappa(\cdot,Y_i)$. Then 
\begin{align*}
    &\|\hat\Sigma\lo{XY}-\tilde\Sigma\lo{XY}\|\lo {\mathrm{HS}}\\
    =&\Bigl\|\frac{1}{n}\sum_{i=1}^n\bigl(\hat{F}_i-\frac{1}{n}\sum_{j=1}^n\hat{F}_j\bigr)\otimes\bigl(\hat{G}_i-\frac{1}{n}\sum_{j=1}^n\hat{G}_j\bigr)-\frac{1}{n}\sum_{i=1}^n\bigl(\tilde{F}_i-\frac{1}{n}\sum_{j=1}^n\tilde{F}_j\bigr)\otimes\bigl(\tilde{G}_i-\frac{1}{n}\sum_{j=1}^n\tilde{G}_j\bigr)\Bigr\|\lo {\mathrm{HS}}\\
    =&\Bigl\|\frac{1}{n}\sum_{i=1}^n\bigl((\hat{F}_i-\tilde{F}_i)-\frac{1}{n}\sum_{j=1}^n(\hat{F}_j-\tilde{F}_j)\bigr)\otimes\bigl((\hat{G}_i-\tilde{G}_i)-\frac{1}{n}\sum_{j=1}^n(\hat{G}_j-\tilde{G}_j)\bigr)\Bigr\|\lo {\mathrm{HS}}\\
    \le&\Bigl\|\frac{1}{n}\sum_{i=1}^n(\hat{F}_i-\tilde{F}_i)\otimes(\hat{G}_i-\tilde{G}_i)-(2-\frac{1}{n})\bigl(\frac{1}{n}\sum_{j=1}^n(\hat{F}_j-\tilde{F}_j)\bigr)\otimes\bigl(\frac{1}{n}\sum_{j=1}^n(\hat{G}_j-\tilde{G}_j)\bigr)\Bigr\|\lo {\mathrm{HS}}\\
    \le&\frac{1}{n}\sum_{i=1}^n\Bigl\|(\hat{F}_i-\tilde{F}_i)\otimes(\hat{G}_i-\tilde{G}_i)\Bigr\|\lo {\mathrm{HS}}+2\Bigl\|\bigl(\frac{1}{n}\sum_{j=1}^n(\hat{F}_j-\tilde{F}_j)\bigr)\otimes\bigl(\frac{1}{n}\sum_{j=1}^n(\hat{G}_j-\tilde{G}_j)\bigr)\Bigr\|\lo {\mathrm{HS}}.\tag{S.2}\label{eq:lemmacov1}
\end{align*}
Consider the expectation of the first term on the right-hand side. Here,  the expectation involves two layers of randomness: that in  $(\{X_{ij}\}^m_{j=1},\{Y_{ik}\}^m_{k=1})$ and that in $X_i$. Taking expectation with respect to $(\{X_{ij}\}^m_{j=1},\{Y_{ik}\}^m_{k=1})$ and then $X \lo i$, we have
\begin{align*}
    E\Bigl[\frac{1}{n}\sum_{i=1}^n\Bigl\|(\hat{F}_i-\tilde{F}_i)\otimes(\hat{G}_i-\tilde{G}_i)\Bigr\|\lo {\mathrm{HS}}\Bigr]&    =\frac{1}{n}\sum_{i=1}^nE\Bigl[\Bigl\|(\hat{F}_i-\tilde{F}_i)\Bigr\|_{\Hcal\lo X}\Bigl\|(\hat{G}_i-\tilde{G}_i)\Bigr\|_{\Hcal\lo Y}\Bigr]\\
    &\le \frac{1}{n}\sum_{i=1}^n\Bigl(E\Bigl\|(\hat{F}_i-\tilde{F}_i)\Bigr\|_{\Hcal\lo X}^2\Bigr)^{1/2}\Bigl(E\Bigl\|(\hat{G}_i-\tilde{G}_i)\Bigr\|_{\Hcal\lo Y}^2\Bigr)^{1/2},
\end{align*}
Evoking the Lipschitz continuity condition on   $\kappa(z,z^\prime)$, we have
\begin{align*}
    E\Bigl\|(\hat{F}_i-\tilde{F}_i)\Bigr\|_{\Hcal\lo X}^2&=E\Bigl\langle(\hat{F}_i-\tilde{F}_i,\hat{F}_i-\tilde{F}_i\Bigr\rangle_{\Hcal\lo X}\\
    &=E\Bigl[\kappa(\hat{X}_i,\hat{X}_i)-2\kappa(\hat{X}_i,X_i)+\kappa(X_i,X_i)\Bigr]\\
    &\le 2CE\Bigl[ d(X_i,\hat{X}_i)\Bigr]\\
    &\le \Ocal \Bigl(E_{X_i}E_{\hat{X}_i}\Bigl[d(X_i,\hat{X}_i)\Bigr]\Bigr).
\end{align*}
By Assumption \ref{assum:emp_true}, $E_{\hat{X}}[d_W(\hat{X_i}, X_i)] =\Ocal(\delta_m)$ for $i=1,\dots,n$. We then have $E\|(\hat{F}_i-\tilde{F}_i)\|_{\Hcal\lo X}^2=\Ocal(\delta_m)$ for $i=1,\dots,n$. Similarly, we have $E\|(\hat{G}_i-\tilde{G}_i)\|_{\Hcal\lo Y}^2=\Ocal(\delta_m)$ for $i=1,\dots,n$. Therefore,
\begin{align}
    E\Bigl[\frac{1}{n}\sum_{i=1}^n\Bigl\|(\hat{F}_i-\tilde{F}_i)\otimes(\hat{G}_i-\tilde{G}_i)\Bigr\|\lo {\mathrm{HS}}\Bigr]=\Ocal(\delta_m).\tag{S.3}\label{eq:lemmacov2}
\end{align}
For the expectation of the second term on the right-hand side of equation \eqref{eq:lemmacov1}, we have
\begin{align*}
    &2E\Bigl\|\bigl(\frac{1}{n}\sum_{j=1}^n(\hat{F}_j-\tilde{F}_j)\bigr)\otimes\bigl(\frac{1}{n}\sum_{j=1}^n(\hat{G}_j-\tilde{G}_j)\bigr)\Bigr\|\lo {\mathrm{HS}}\\
    =&2E\Bigl[\Bigl\|\bigl(\frac{1}{n}\sum_{j=1}^n(\hat{F}_j-\tilde{F}_j)\bigr)\Bigr\|_{\Hcal\lo X}\Bigl\|\bigl(\frac{1}{n}\sum_{j=1}^n(\hat{G}_j-\tilde{G}_j)\bigr)\Bigr\|_{\Hcal\lo Y}\Bigr]\\
    \le&2\Bigl(E\Bigl\|\frac{1}{n}\sum_{j=1}^n(\hat{F}_j-\tilde{F}_j)\Bigr\|_{\Hcal\lo X}^2\Bigr)^{1/2}\Bigl(E\Bigl\|\frac{1}{n}\sum_{j=1}^n(\hat{G}_j-\tilde{G}_j)\Bigr\|^2_{\Hcal\lo Y}\Bigr)^{1/2}\\
    \le&2\Bigl(\frac{1}{n}\sup_{1\le i\le n}E\Bigl\|(\hat{F}_i-\tilde{F}_i)\Bigr\|_{\Hcal\lo X}^2\Bigr)^{1/2}\Bigl(\frac{1}{n}\sup_{1\le i\le n}E\Bigl\|(\hat{G}_i-\tilde{G}_i)\Bigr\|^2_{\Hcal\lo Y}\Bigr)^{1/2}\\
    \le& \Ocal(\delta_m/n).\tag{S.4}\label{eq:lemmacov3}
\end{align*}
Combine result \eqref{eq:lemmacov0}\eqref{eq:lemmacov2} and \eqref{eq:lemmacov3}, we have 
\begin{align*}
    E\|\hat\Sigma\lo{XY}-\Sigma\lo{XY}\|\lo {\mathrm{HS}}\le\Ocal(\delta_m(1+1/n)+n^{-1/2})=\Ocal(\delta\lo m + n\hi{-1/2}). 
\end{align*}
Then by Chebyshev’s inequality, we have
\begin{align*}
    \|\hat\Sigma\lo{XY}-\Sigma\lo{XY}\|\lo {\mathrm{HS}}\le\Ocal\lo p(\delta\lo m + n\hi{-1/2}), 
\end{align*}
as desired. This completes the proof  of Lemma \ref{lemma:cov}.
\eop
\end{proof}

\subsection{Proof of Theorem \ref{thm: disobs}}
\begin{proof}
Let
\begin{align*}
    \hat{A} = (\hat\Sigma\lo{XX} + \eta_nI)\inv,\quad A_n=(\Sigma\lo{XX} + \eta_nI)\inv,\quad A=\Sigma\lo{XX}\inv;\quad \hat{B}= \hat\Sigma\lo{XY},\quad B=\Sigma\lo{XY}.
\end{align*}
Then the element of interest $\hat\Lambda-\Lambda$ can be writen as
\begin{align*}
    \hat\Lambda-\Lambda&=\hA\hB\hB^*\hA^*-AB\sB\sA\\
    &=\hA\hB(\hB^*\hA^*- B^*A^*) + ( \hA\hB-AB)B^*A^*;
\end{align*}
Thus, we have
\begin{align*}
    \|\hat\Lambda-\Lambda\|_{\mathrm{OP}}&\le\|\hA\hB(\hB^*\hA^*- B^*A^*)\|_{\mathrm{OP}} + \|(\hA\hB-AB)B^*A^*\|_{\mathrm{OP}}\\
    &=\|(AB-\hA\hB)\hB^*\hA^*\|_{\mathrm{OP}} + \|(\hA\hB-AB)B^*A^*\|_{\mathrm{OP}}\\
    &\le \|(AB-\hA\hB)\|_{\mathrm{OP}}(\|\hA\hB\|_{\mathrm{OP}}+\|AB\|_{\mathrm{OP}}). 
\end{align*}
Since both $AB$ and $\hA\hB$ are compact operators, it suffices to show that
\begin{align*}
    \|(AB-\hA\hB)\|_{\mathrm{OP}}=\Ocal_p(\eta_n^\beta+\eta_n\inv\varepsilon_{n,m}),
\end{align*}
where $\varepsilon\lo{n, m} = \delta_m+n^{-1/2}$.
Writing $\hA\hB$ as
\begin{align*}
    \hA\hB=\hA(\hB- B) + (\hA- A_n)B + (A_n- A)B + AB,
\end{align*}
we obtain
\begin{align}\label{eq:thm1}
    \|(AB-\hA\hB)\|_{\mathrm{OP}}\le\|\hA(\hB- B)\|_{\mathrm{OP}} + \|(\hA- A_n)B\|_{\mathrm{OP}} + \|(A_n- A)B\|_{\mathrm{OP}}. \tag{S.5}
\end{align}
For the first term on the right-hand side, we have
\begin{align*}
    \|\hA(\hB- B)\|_{\mathrm{OP}}=&\|(\hat{\Sigma}_{XX}+\eta_nI)\inv(\hat{\Sigma}_{XY}- \Sigma\lo{XY})\|_{\mathrm{OP}}\\
    \le&\|(\hat{\Sigma}_{XX}+\eta_nI)\inv\|_{\mathrm{OP}}\|(\hat{\Sigma}_{XY}- \Sigma\lo{XY})\|_{\mathrm{HS}}\\
    \le&\eta_n\inv\|(\hat{\Sigma}_{XX}+\eta_nI)(\hat{\Sigma}_{XX}+\eta_nI)\inv\|_{\mathrm{OP}}\|(\hat{\Sigma}_{XY}- \Sigma\lo{XY})\|_{\mathrm{HS}}\\
    \le&\Ocal_p(\eta_n\inv\varepsilon_{n,m}),\tag{S.6} \label{eq:thm2}
\end{align*}
where the last inequality follows from Lemma \ref{lemma:cov}.
For the second term on the right-hand side of \eqref{eq:thm1}, we write it as
\begin{align*}
    (\hA- A_n)B=&\bigl((\hat{\Sigma}_{XX}+\eta_nI)\inv-(\Sigma\lo{XX} + \eta_nI)\inv\bigr)\Sigma\lo{XY}\\
    =&(\hat{\Sigma}_{XX}+\eta_nI)\inv\bigl((\hat{\Sigma}_{XX}+\eta_nI)-(\Sigma\lo{XX} + \eta_nI)\bigr)(\Sigma\lo{XX} + \eta_nI)\inv\Sigma\lo{XY}\\
    =&(\hat{\Sigma}_{XX}+\eta_nI)\inv(\hat{\Sigma}_{XX}-\Sigma\lo{XX})(\Sigma\lo{XX} + \eta_nI)\inv\Sigma\lo{XX}\Sigma\lo{XX}\inv\Sigma\lo{XY}.
\end{align*}
Thus, we have
\begin{align*}
    &\|(\hA- A_n)B\|_{\mathrm{OP}}\\
    \le& \|(\hat{\Sigma}_{XX}+\eta_nI)\inv\|_{\mathrm{OP}}\|\hat{\Sigma}_{XX}-\Sigma\lo{XX}\|_{\mathrm{OP}}\|(\Sigma\lo{XX} + \eta_nI)\inv\Sigma\lo{XX}\|_{\mathrm{OP}}\|\Sigma\lo{XX}\inv\Sigma\lo{XY}\|_{\mathrm{OP}}. 
\end{align*}
By the above derivations, we have  $\|(\hat{\Sigma}_{XX}+\eta_nI)\inv\|_{\mathrm{OP}}=\Ocal_p(\eta_n \inv )$ and $\|\hat{\Sigma}_{XX}-\Sigma\lo{XX}\|_{\mathrm{OP}}=\Ocal_p(\varepsilon_{n,m})$. Also, we have
\begin{align*}
    \|(\Sigma\lo{XX} + \eta_nI)\inv\Sigma\lo{XX}\|_{\mathrm{OP}}\le \|(\Sigma\lo{XX} + \eta_nI)\inv(\Sigma\lo{XX}+\eta_nI)\|_{\mathrm{OP}}=1,
\end{align*}
and $\|\Sigma\lo{XX}\inv\Sigma\lo{XY}\|_{\mathrm{OP}}\le \infty$ by Assumption \ref{assum:ker}. Therefore, we have
\begin{align}\label{eq:thm3}
    \|(\hA- A_n)B\|_{\mathrm{OP}}=\Ocal_p(\eta_n\inv\varepsilon_{n,m}).\tag{S.7}
\end{align}
Finally, letting $R_{XY}=\Sigma\lo{XY}^{\beta}S_{XY}$ and rewriting the third term on the right-hand side  of \eqref{eq:thm1} as
\begin{align*}
    (A_n- A)B=&((\Sigma\lo{XX} + \eta_nI)\inv-\Sigma\lo{XX}\inv)\Sigma\lo{XY}\\
    =&(\Sigma\lo{XX} + \eta_nI)\inv\Sigma\lo{XX}R_{XY}-R_{XY}\\
    =&-\eta_n(\Sigma\lo{XX} + \eta_nI)\inv R_{XY}, 
\end{align*}
we see that 
\begin{align*}
    \|(A_n- A)B\|_{\mathrm{OP}}&\le \eta_n\|(\Sigma\lo{XX} + \eta_nI)^{-1+\beta}\|_{\mathrm{OP}}\|S_{XY}\|_{\mathrm{OP}}\\
    &\le \eta_n\eta_n^{\beta-1}\|(\Sigma\lo{XX} + \eta_nI)^{-1+\beta}(\Sigma\lo{XX} + \eta_nI)^{1-\beta}\|_{\mathrm{OP}}\|S_{XY}\|_{\mathrm{OP}}\\
    &\le\eta_n^\beta\|S_{XY}\|_{\mathrm{OP}}=\Ocal_p(\eta_n^\beta). \tag{S.8}\label{eq:thm4}
\end{align*}
Combining   \eqref{eq:thm2},  \eqref{eq:thm3},  and \eqref{eq:thm4}, we prove the first assertion of the theorem. 

The second assertion can then be proved by following roughly the same path and using the following facts:

\begin{enumerate}
\item[(i)] if $A$ is a bounded operator and $B$ is Hilbert-Schmidt operator and $\ran(A)\subseteq\mathrm{dom}(B)$, then $AB$ is a Hilbert Schmidt operator with $$\|AB\|_{\mathrm{HS}} \le \|A\|_{\mathrm{OP}}\|B\|_{\mathrm{HS}};$$
\item[(ii)]  if $A$ is
Hilbert-Schmidt then so is $A^*$ and $$\|A\|_{\mathrm{HS}} = \|A^*\|_{\mathrm{HS}}.$$
\end{enumerate}

Using the same decomposition as \eqref{eq:thm1}, we have
\begin{align}\label{eq:thm5}
    \|(AB-\hA\hB)\|_{\mathrm{HS}}\le\|\hA(\hB- B)\|_{\mathrm{HS}} + \|(\hA- A_n)B\|_{\mathrm{HS}} + \|(A_n- A)B\|_{\mathrm{HS}}.\tag{S.9}
\end{align}
For the first term on the right-hand side of (\ref{eq:thm5}):
\begin{align}\label{eq:thm6}
    \|\hA(\hB- B)\|_{\mathrm{HS}}\le\|\hA\|_{\mathrm{OP}}\|\hB- B\|_{\mathrm{HS}}=\Ocal(\eta_n\inv\varepsilon_{n,m}).\tag{S.10}
\end{align}
For the second term on the right-hand side of (\ref{eq:thm5}):
\begin{align*}
    &\|(\hA- A_n)B\|_{\mathrm{HS}}\\
    \le& \|(\hat{\Sigma}_{XX}+\eta_nI)\inv\|_{\mathrm{OP}}\|\hat{\Sigma}_{XX}-\Sigma\lo{XX}\|_{\mathrm{OP}}\|(\Sigma\lo{XX} + \eta_nI)\inv\Sigma\lo{XX}\|_{\mathrm{OP}}\|\Sigma\lo{XX}\inv\Sigma\lo{XY}\|_{\mathrm{HS}}\\
    \le& \|(\hat{\Sigma}_{XX}+\eta_nI)\inv\|_{\mathrm{OP}}\|\hat{\Sigma}_{XX}-\Sigma\lo{XX}\|_{\mathrm{OP}}\|(\Sigma\lo{XX} + \eta_nI)\inv\Sigma\lo{XX}\|_{\mathrm{OP}}\|\Sigma\lo{XX}^{\beta}S_{XY}\|_{\mathrm{HS}}\\
    =&\Ocal_p(\eta_n\inv\varepsilon_{n,m}).\tag{S.11}\label{eq:thm7}
\end{align*}
For the third term on the right-hand side of (\ref{eq:thm5}):
\begin{align}\label{eq:thm8}
    \|(A_n- A)B\|_{\mathrm{HS}}\le \eta_n\|(\Sigma\lo{XX} + \eta_nI)^{-1+\beta}\|_{\mathrm{OP}}\|S_{XY}\|_{\mathrm{HS}}\le\eta_n^\beta\|S_{XY}\|_{\mathrm{HS}}=\Ocal_p(\eta_n^\beta).\tag{S.12}
\end{align}
Combining the results \eqref{eq:thm6}, \eqref{eq:thm7},  and \eqref{eq:thm8}, we have 
\begin{align*}
    \|(AB-\hA\hB)\|_{\mathrm{HS}}\le\Ocal_p(\eta_n^\beta+\eta_n\inv\varepsilon_{n,m}).
\end{align*}
This completes the proof  of Theorem \ref{thm: disobs}. \eop 
\end{proof}

\bibliographystyle{agsm}
\bibliography{ref.bib}


\end{document}